%% file: RareTransition_journal.tex
\documentclass[english,journal,11pt,draftcls,onecolumn]{IEEEtran}
\IEEEoverridecommandlockouts

\usepackage{amssymb}
\usepackage[cmex10]{amsmath}
\usepackage{bm}
\usepackage{verbatim}
\usepackage{graphicx}
\usepackage{tikz,pgf,pgfplots}
\usepackage{units}
\usepgflibrary{shapes}
\usetikzlibrary{%
  arrows,%
  shapes,%
  chains,%
  matrix,%
  positioning,% wg. " of "
  shadows,%
  fit,%
  petri%
}

\newtheorem{lemma}{Lemma}
\newtheorem{theorem}{Theorem}
\newtheorem{remark}{Remark}
\newtheorem{proposition}{Proposition}
\newtheorem{corollary}{Corollary}
\newtheorem{definition}{Definition}
\newtheorem{assumption}{Assumption}

\newcommand{\bad}{\ensuremath{\mathrm{b}}}
\newcommand{\good}{\ensuremath{\mathrm{g}}}
\newcommand{\type}{\ensuremath{\mathcal{T}}}

\begin{document}

\title{On the Performance of Short Block Codes\\ over Finite-State Channels in the\\ Rare-Transition Regime}

\author{\IEEEauthorblockN{Fatemeh Hamidi-Sepehr, %\emph{Student Member, IEEE},
Jean-Francois Chamberland, %\emph{Senior Member, IEEE},
Henry D. Pfister%, \emph{Senior Member, IEEE}
}
\thanks{This material is based upon work supported by the National Science Foundation (NSF) under Grants No.~0747363 and No.~0830696.
Any opinions, findings, and conclusions or recommendations expressed in this material are those of the authors and do not necessarily reflect the views of the National Science Foundation.
This paper was presented in part at the 50th Annual Allerton Conference on Communication, Control, and Computing and the 46th Annual Conference on Information Sciences and Systems. 

The authors are with the Department of Electrical and Computer Engineering, Texas A\&M University, College Station, TX 77843, USA (emails: f\_hamidisepehr@tamu.edu; chmbrlnd@tamu.edu; hpfister@tamu.edu).}
}

\maketitle

\begin{abstract}
As the mobile application landscape expands, wireless networks are tasked with supporting different connection profiles, including real-time traffic and delay-sensitive communications.
Among many ensuing engineering challenges is the need to better understand the fundamental limits of forward error correction in non-asymptotic regimes.
This article characterizes the performance of random block codes over finite-state channels and evaluates their queueing performance under maximum-likelihood decoding.
In particular, classical results from information theory are revisited in the context of channels with rare transitions, and bounds on the probabilities of decoding failure are derived for random codes.
This creates an analysis framework where channel dependencies within and across codewords are preserved.
Such results are subsequently integrated into a queueing problem formulation.
For instance, it is shown that, for random coding on the Gilbert-Elliott channel, the performance analysis based on upper bounds on error probability provides very good estimates of system performance and optimum code parameters.
Overall, this study offers new insights about the impact of channel correlation on the performance of delay-aware, point-to-point communication links.
It also provides novel guidelines on how to select code rates and block lengths for real-time traffic over wireless communication infrastructures.
\end{abstract}

\section{Introduction}
\label{section:Introduction}

With the ever increasing popularity of advanced mobile devices such as smartphones and tablet personal computers, the demand for ubiquitous low-latency, high-throughput wireless services is growing rapidly.
The shared desire for a heightened user experience, which includes real-time applications and mobile interactive sessions, acts as a motivation for the study of highly efficient communication schemes subject to stringent delay constraints.
An important aspect of delay-sensitive traffic stems from the fact that intrinsic delivery requirements may preclude the use of very long codewords.
As such, the insights offered by classical information theory and based on Shannon capacity are of limited value in this context.
A prime goal of this article is to develop a better understanding of delay-constrained communication, queue-based performance criteria and service dependencies attributable to channel memory.
In particular, we seek to derive meaningful performance limits for delay-aware systems operating over channels with memory.
The emphasis is put on identifying upper bounds on the probabilities of decoding failure for systems employing short block lengths.
This is an essential intermediate step in the characterization of queueing behavior for contemporary communication systems.

Computing the probability of decoding failure for specific communication channels and particular coding schemes is of great interest.
This line of work dates back to the early days of information theory~\cite{Fano-1961} and has received significant attention in the past, with complete solutions in some cases.
One approach that has been remarkably successful, consists of deriving exponential error bounds on the behavior of asymptotically long codewords.
This approach was popularized by Gallager~\cite{Gallager-1968} and these bounding techniques have been applied to both memoryless channels and finite-state channels with memory.
In general, they are very accurate for long, yet finite block lengths.
It is worth mentioning that the subject of error bounds has also appeared in many more recent studies, with the advent of new approaches such as dispersion, the uncertainty-focusing bound, and the saddlepoint approximation~\cite{Barg-it02,Sahai-it08,Polyanskiy-it10,Polyanskiy-it11,Martinez-ITA11}.
This renewed interest in the performance of coded transmissions points to the timeliness of the topic under investigation.

A distinguishing feature of the approach that we wish to develop is the focus on indecomposable channels with memory and state-dependent operation.
%It is a well-known fact that the concept of channel indecomposability coincides with that of the ergodicity of underlying Markov  process.
In many asymptotic frameworks, channel parameters are kept constant while the length of the codeword increases to infinity.
Although this point of view leads to mathematically appealing characterizations, the resulting bounds on error probabilities do not depend on the initial or final states of the underlying communication channel.
This situation can be explained through the fact that, no matter how slow the mixing time of the physical channel is, the duration of a codeword eventually far exceeds this quantity.
Still, in many practical scenarios, the service requirements imposed on a communication link forces the use of relatively short codewords, with no obvious time-scale separation between the duration of a codeword and the mixing time of the channel.
In practice, the tradeoff between performance and delay encourages system designers to choose block lengths of the same order as the channel mixing time.

The mismatch between existing techniques and commonly deployed systems, together with the growing popularity of real-time applications on wireless networks, demands a novel approach where the impact of boundary conditions are preserved throughout the analysis.
A suitable methodology should be able to capture both the effects of channel memory as well as the impact of the channel state at the onset of a codeword.
In this article, we are interested in regimes where the block length is of the same order or smaller than the coherence time of the channel.
Formally, we wish to study the scenario where the mixing time of the underlying finite-state channel is similar to the time necessary to transmit a codeword.
This leads to two important phenomena.
First, the state of the channel at the onset of a transmission has a significant impact on the empirical distribution of the states within a codeword transmission cycle.
Second, channel dependencies may extend beyond the boundaries of individual codewords.
This is in stark contrast with rapidly mixing channels where initial channel conditions  have no effects on the probability of decoding failures.
It also differs from block-fading models where the evolution of the channel is independent from block to block.
Our proposed framework is rich enough to account for scenarios where decoding events are strongly correlated over time.
Dependencies from codeword to codeword give rise to correlation in decoding failure events and can therefore greatly affect perceived service quality.

It is important to note that the standard asymptotic regimes naturally result in concentration of channel quality over a codeword transmission time.
This enables the application of the law of large numbers to compute the mutual information, the convergence rate, etc.
However, in non-asymptotic analysis of channels with memory, variations of the channel quality over a codeword transmission time need careful consideration.
At the same time, high correlations make the channel behavior more predictable.
This in turn, paves the way to implement adaptive opportunistic scheduling schemes.
Our proposed methodology focuses on the rare-transition regime and provides results that are asymptotic in the number of bits but not in the elapsed time.

It is worth pointing out that the rare-transition regime, also known as the slow-mixing regime, has been studied in other contexts such as channel estimation, asymptotic filtering, and entropy rate analysis of hidden Markov processes \cite{Nair-isit05,Peres-arXiv10,Asadi-arXiv13,Pinsker-00}.
In this article, we examine probabilities of decoding failure, their distributions and temporal properties within the context of rare-transition regime.
The purpose of deriving upper bounds on the probabilities of decoding failure for rare transitions is to capture overall performance for systems that transmit data using block lengths on the order of the coherence time of their respective channels.
More specifically, this article focuses on Gallager-type exponential bounds applied to probabilities of decoding failure in rare-transition regimes.
By construction, these bounds depend explicitly on the initial and terminating channel states at the codeword boundaries.
The analysis is conducted for the scenario where channel state information is available at the receiver.
The ensuing results are subsequently compared to the probabilities of decoding failure obtained for a Gilbert-Elliott channel under a minimum distance decoder and a maximum-likelihood decision rule \cite{Fano-1961,Gilbert-bell60,Elliott-bell63}.

The potential implications of this novel framework are then discussed in terms of queueing theory.
We exploit the results of the error-probability analysis in the rare-transition regime to evaluate the queueing performance of correlated channels.
Particularly, we employ the derived upper bounds on the probability of decoding failure to bound the queueing performance of the system.
In that regard, we show how stochastic dominance enables a tractable analysis of overall performance.
The results of this evaluation are then compared with a performance characterization based on the exact probability of decoding failure for a Gilbert-Elliott channel under maximum-likelihood decoding.

\section{Modeling and Exponential Bounds}
\label{section:Modeling}

We consider indecomposable finite-state channels where state transitions are independent of the input symbols.
Such channels are often classified as fading models, and they have been used extensively in the information theory literature.
We employ $X_n$ and $Y_n$, respectively, to denote the input and output symbols at time~$n$.
The channel state that determines the channel law at time~$n$ is represented by $S_{n-1}$.
We typically reserve capital letters for random variables, whereas lower case letters identify outcomes and values.
Boldface letters are used to denote length-$N$ sequences of random variables or outcomes.
For groups of random variables, we use the common expression $P _{\cdot|\cdot} (\cdot|\cdot)$ to denote the conditional joint probability mass function, and $P _{\mathrm{e},\cdot|\cdot} (\cdot|\cdot)$ to denote the conditional joint probability of decoding error.

In general, the conditional probability distribution governing a finite-state channel can be written as
\begin{equation*}
\begin{split}
&P_{Y_n,S_n | X_n,S_{n-1}} \left( y_n, s_n | x_n, s_{n-1} \right) \\
&= \Pr \left( Y_n = y_n, S_n = s_n | X_n = x_n, S_{n-1} = s_{n-1} \right) .
\end{split}
\end{equation*}
When state transitions are independent of input symbols, this conditional distribution allows the \linebreak {factorization}
\begin{equation} \label{equation:ChannelDecomposition}
\begin{split}
&P_{Y_n, S_n | X_n, S_{n-1}} ( y_n, s_n | x_n, s_{n-1} ) \\
&= P_{S_n | S_{n-1}} ( s_n | s_{n-1} )
P_{Y_n | X_n, S_{n-1}} ( y_n | x_n, s_{n-1} ) .
\end{split}
\end{equation}
Throughout, we assume that the channel statistics are homogeneous over time and the sequence $\{ S_n \}$ forms a Markov chain.
When dealing with finite-state channels, it is customary to use integers to denote the possible input and output symbols; in our exposition, we adhere to this convention.

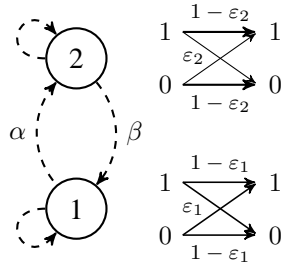
\begin{figure}[tb]
\centering
\begin{tikzpicture}
[node distance = 10mm, draw=black, thick, >=stealth',
state/.style={circle, draw=black, inner sep = 0pt, minimum size = 8mm}]

\node[state] (l0) at (0,0) {1};
\draw [->, dashed] (l0) to [in=230, out=180, looseness=5] (l0);

\node[state] (l1) at (0,2) {2}
  edge[<-, dashed, bend right=40] node[left] {\small{${\alpha}$}} (l0)
  edge[->, dashed, bend left=40] node[right] {\small{${\beta}$}} (l0);
\draw [->, dashed] (l1) to [in=130, out=180, looseness=5] (l1);

\node[coordinate] (l0In0) [right=of l0, yshift=-10, label=left:{\small{$0$}}]{};
\node[coordinate] (l0In1) [right=of l0, yshift=10, label=left:{\small{$1$}}]{};

\node[coordinate] (l0Out0) [right= of l0In0, label=right:{\small{$0$}}]{}
  edge[pre, semithick] node[below] {\scriptsize{$1 - \varepsilon_1$}} (l0In0)
  edge[pre, semithick] node[left] {\scriptsize{$\varepsilon_1$}\;} (l0In1);

\node[coordinate] (l0Out1) [right= of l0In1, label=right:{\small{$1$}}]{}
  edge[pre, semithick] node[above] {\scriptsize{$1 - \varepsilon_1$}} (l0In1)
  edge[pre, semithick] (l0In0);

\node[coordinate] (l1In0) [right=of l1, yshift=-10, label=left:{\small{$0$}}]{};
\node[coordinate] (l1In1) [right=of l1, yshift=10, label=left:{\small{$1$}}]{};

\node[coordinate] (l1Out0) [right= of l1In0, label=right:{\small{$0$}}]{}
  edge[pre, thick] node[below] {\scriptsize{$1 - \varepsilon_2$}} (l1In0)
  edge[pre, thin] node[left] {\scriptsize{$\varepsilon_2$}\;} (l1In1);

\node[coordinate] (l0Out1) [right= of l1In1, label=right:{\small{$1$}}]{}
  edge[pre, thick] node[above] {\scriptsize{$1 - \varepsilon_2$}} (l1In1)
  edge[pre, thin] (l1In0);
\end{tikzpicture}
\caption{The Gilbert-Elliott model is the simplest, non-trivial instantiation of a finite-state channel with memory.
State evolution over time forms a Markov chain and the input-output relationship of this binary channel is governed by a state-dependent crossover probability.}
\label{figure:GilbertElliott}
\end{figure}

The famed Gilbert-Elliott channel is the proverbial example of a channel that possesses the structure described above \cite{Gilbert-bell60,Elliott-bell63}.
This quintessential model is governed by a two-state Markov chain, and it is illustrated in Fig.~\ref{figure:GilbertElliott}.
The transition probability matrix for the Gilbert-Elliott channel can be expressed as
\begin{equation} \label{equation:ProbabilityTransitionMatrix}
\mathbf{P} = \begin{bmatrix}
1 - {\alpha} & {\alpha} \\
{\beta} & 1 - {\beta}
\end{bmatrix} ,
\end{equation}
where $\left[ \mathbf{P} \right]_{ij} = \Pr (S_n = j | S_{n-1} = i)$.
The state-dependent input-output relationship induced by channel state $s \in \{ 1, 2 \}$ is governed by the crossover probability $\varepsilon_s$, where
\begin{align*}
\Pr ( x_n = y_n | S_{n-1} = s ) &= 1 - \varepsilon_s \\
\Pr ( x_n \neq y_n | S_{n-1} = s ) &= \varepsilon_s .
\end{align*}
For convenience, we order states such that $\varepsilon_1 < \varepsilon_2$.

A coding strategy that has proven exceptionally fruitful in information theory is the use of random codes.
Continuing this tradition, we adopt a random coding scheme that employs a code ensemble $\mathcal{C}$ with $M = e^{NR}$ elements.
Familiarity with this topic may help because random coding arguments tend to be notationally heavy~\cite{Gallager-1968}.
The variable $N$ denotes the block length of the code and $R$ is the code rate in nats per code bit.
Every element in $\mathcal{C}$ corresponds to a sequence of admissible channel inputs,
$\mathbf{x} = (x_1, x_2, \ldots, x_N)$.
Moreover, codewords are indexed by $k \in \{ 1, \ldots, M \}$.
The input sequence associated with index~$k$, which we denote by $\mathbf{X}(k)$, is determined through the following procedure.
Suppose that $Q(\cdot)$ is a distribution on the set of admissible input symbols.
Let $\mathbf{Q}_N (\mathbf{x}) = \prod_{n=1}^N Q(x_n)$ be the product measure induced by $Q$.
Codeword $\mathbf{X}(k)$ is selected at random according to distribution $\mathbf{Q}_N$, i.e.,
\begin{equation*}
\Pr (\mathbf{X}(k) = \mathbf{x}) = \mathbf{Q}_N (\mathbf{x})
= \prod_{n=1}^N Q(x_n) .
\end{equation*}
We emphasize that every codeword is selected independently from other elements in $\mathcal{C}$.
Once a code ensemble has been generated, a message is sent to the destination by first selecting one of the codewords, and then sequentially transmitting its entries over the communication channel.
For the sake of clarity, we summarize our assumptions below; they apply from this point forward, unless otherwise stated.

\begin{assumption} \label{assumption:DecompositionRandomCodes}
Communication takes place over a finite-state channel that admits the conditional decomposition of \eqref{equation:ChannelDecomposition}.
Information is transmitted using the random coding strategy outlined above.
On the receiver side, a maximum-likelihood decision rule is used to decode the received sequence.
Furthermore, the state of the channel is causally known at the receiver.
\end{assumption}

At this point, it is worth restating our objective.
We want to upper bound the probability that a codeword is decoded erroneously at the receiver.
Concurrently, we wish to develop a rare-transition regime that remains true to the fact that the channel state at the onset of the codeword transmission process affects the evolution of the system.
Ultimately, this can be achieved in an asymptotic framework by slowing down the transition profile of the underlying channel as the block length of the code grows unbounded.
One of the repercussions of this setting is that we have to modify some of the results on error exponents presented by Gallager~\cite{Gallager-1968}.
In particular, we need the ability to restrict a channel sequence $\mathbf{S}$ to specific events and decompose the error probability accordingly.

Our first formal result is a straightforward extension to Theorem~5.6.1 in \cite[pp.~135]{Gallager-1968}, which is itself quite general.
Since we are interested in finite-state channels with memory in a slow transition regime, we require the ability to track channel realizations explicitly.
From an abstract perspective, conditioning on a specific fading realization is equivalent to altering the statistical profile of the underlying channel.

\begin{proposition} \label{proposition:UpperBound}
Suppose that the realization of the channel over the duration of a codeword is given by $\mathbf{s}$.
Then, for any $\rho \in [0,1]$, the probability of decoding failure at the destination, conditioned on state sequence $\mathbf{S} = \mathbf{s}$, is upper bounded by
\begin{equation*}
P_{\mathrm{e} | \mathbf{S}} (\mathbf{s})
\leq e^{- N (E_{0,N} (\rho, \mathbf{Q}_N, \mathbf{s}) - \rho R) }
\end{equation*}
where the exponent $E_{0,N} (\rho, \mathbf{Q}_N, \mathbf{s})$ is equal to
\begin{equation*}
- \frac{1}{N} \ln \! \prod_{n=1}^N \sum_{y_n} \! \left[ \sum_{x_n} Q(x_n)
P_{Y_n | X_n, S_{n \! - \! 1}} ( y_n | x_n, s_{n \! - \! 1} )^{\frac{1}{1 + \rho}}
\right]^{1 + \rho} \! .
\end{equation*}
\end{proposition}
\begin{IEEEproof}
First, we emphasize that the condition $\mathbf{S} = \mathbf{s}$ simply alters the probability measure governing the input-output relationship of the channel.
Applying Theorem~5.6.1 in~\cite{Gallager-1968} with $M = e^{NR}$, we immediately get
\begin{equation*}
P_{\mathrm{e} | \mathbf{S}} (\mathbf{s})
\leq e^{\rho NR} \sum_{\mathbf{y}} \! \left[ \sum_{\mathbf{x}}
\mathbf{Q}_N  (\mathbf{x})
P_{\mathbf{Y}|\mathbf{X},\mathbf{S}} \left( \mathbf{y} | \mathbf{x},\mathbf{s} \right)^{\frac{1}{1+\rho}} \right]^{1 + \rho} ,
\end{equation*}
where $P_{\mathbf{Y}|\mathbf{X},\mathbf{S}} ( \mathbf{y} | \mathbf{x},\mathbf{s} )$ represents the conditional distribution of receiving $\mathbf{y}$ given $\mathbf{X} = \mathbf{x}$ and $\mathbf{S} = \mathbf{s}$.
Moving forward, the crux of the argument is based on interchanging the order of exhaustive products and sums.
Under the channel decomposition introduced in \eqref{equation:ChannelDecomposition}, the double summation that appears in this upper bound becomes
\begin{equation*}
\begin{split}
&\sum_{\mathbf{y}} \left[ \sum_{\mathbf{x}}
\prod_{n=1}^N Q  (x_n) P_{Y_n | X_n, S_{n - 1}} ( y_n | x_n, s_{n - 1} )^{\frac{1}{1+\rho}} \right]^{1 + \rho} \\
&= \prod_{n=1}^N \sum_{y_n} \left[ \sum_{x_n}
Q  (x_n) P_{Y_n | X_n, S_{n - 1}} ( y_n | x_n, s_{n - 1} )^{\frac{1}{1+\rho}} \right]^{1 + \rho} .
\end{split}
\end{equation*}
Collecting these various results and using equivalent notation, we obtain the desired proposition.
\end{IEEEproof}

A key insight revealed through the proof of Proposition~\ref{proposition:UpperBound} is that $E_{0,N} (\rho, \mathbf{Q}_N, \mathbf{s})$ only depends on $\mathbf{s}$ through its empirical distribution, designated $\mathcal{T} (\mathbf{s})$.
We state this fact as a corollary because it will become very useful shortly.

\begin{corollary} \label{corollary:UpperBound}
Let $T$ be the empirical state distribution of a sequence of $N$ consecutive channel realizations.
If $\mathbf{s}$ and $\mathbf{s}'$ are two sequences such that $\mathcal{T} (\mathbf{s}) = \mathcal{T} (\mathbf{s}') = T$, then with some abuse of notation we can write
\begin{equation*}
E_{0,N} (\rho, \mathbf{Q}_N, \mathbf{s})
= E_{0,N} (\rho, \mathbf{Q}_N, \mathbf{s}')
\triangleq E_{0,N} (\rho, \mathbf{Q}_N, T).
\end{equation*}
%Note that,  we have implicitly defined
%$E_{0,N} (\rho, \mathbf{Q}_N, T) = E_{0,N} (\rho, \mathbf{Q}_N, \mathbf{s})$. % where $\mathbf{s}$ represents any sequence with empirical distribution $\mathcal{T}(\mathbf{s}) = T$.
Furthermore, the probability of decoding failure at the destination, conditioned on $\mathbf{S} =\mathbf{s}$, is bounded by
\begin{equation*}
P_{\mathrm{e}|\mathbf{S}} (\mathbf{s})
\leq e^{- N (E_{0,N} (\rho, \mathbf{Q}_N, T) - \rho R )},
\end{equation*}
for any sequence $ \mathbf{s}$ with empirical distribution $ \mathcal{T} (\mathbf{s}) = T $ and $\rho \in [0,1]$.
\end{corollary}

For the problem at hand, we are especially interested in probabilities of the form $P_{\mathrm{e}, S_N | S_0}(s_N | s_0)$.
In some sense, each of these represents the probability of a decoding failure while keeping track of boundary states.
In view of Corollary~\ref{corollary:UpperBound}, it is natural to upper bound this quantity by partitioning the set of possible sequences according to their empirical distributions.
This is accomplished below.
In stating our results, we use $\mathcal{T}$ to denote the collection of all admissible empirical channel distributions over sequences of length $N$.

\begin{proposition} \label{theorem:EmpiricalDistributionBound}
Suppose that a codeword is transmitted over a finite-state channel.
The joint probability that decoding fails at the destination and $S_N = s_N$, conditioned on initial state $S_0 = s_0$, is upper bounded as follows
\begin{equation} \label{eq:mainbound}
\begin{split}
% SINGLE
P_{\mathrm{e}, S_N|S_0} (s_N|s_0)
&\leq\sum_{T \in \mathcal{T}}
P_{\type (\mathbf{S}), S_N | S_0} (T, s_N | s_0)
\min_{\rho\in[0,1]}  e^{-N (E_{0,N} (\rho, \mathbf{Q}_N, T) - \rho R)} \\
&\leq \min_{\rho \in [0,1]} \sum_{T \in \mathcal{T}}
P_{\type (\mathbf{S}), S_N | S_0} (T, s_N | s_0)
e^{-N (E_{0,N} (\rho, \mathbf{Q}_N, T) - \rho R)}
\end{split}
\end{equation}
where $P_{\type (\mathbf{S}), S_N | S_0} (T, s_N | s_0)$ represents the probability that $\mathcal{T}(\mathbf{S}) = T$ and $S_N = s_N$, given initial state $S_0 = s_0$.
\end{proposition}
\begin{IEEEproof}[Proof]
The progression of this demonstration parallels an argument found in Section~5.9 of \cite{Gallager-1968}.
By partitioning the set of length-$N$ sequences according to their empirical distributions, we can write
\begin{equation*}
\begin{split}
&P_{\mathrm{e},S_N|S_0} (s_N|s_0)
= \sum_{\mathbf{s}} \Pr ( \mathbf{S} = \mathbf{s}, S_N = s_N | S_0 = s_0)
P_{\mathrm{e} | \mathbf{S}} (\mathbf{s}) \\
&\qquad= \sum_{T \in \type} \sum_{\mathbf{s} : \type (\mathbf{s}) = T }
\Pr (\mathbf{S} = \mathbf{s}, S_N = s_N | S_0 = s_0)
P_{\mathrm{e} | \mathbf{S}} (\mathbf{s}) \\
&\qquad\leq \sum_{T \in \type} \sum_{\mathbf{s} : \type (\mathbf{s}) = T }
\Pr (\mathbf{S} = \mathbf{s}, S_N = s_N | S_0 = s_0)
% SINGLE
\min_{\rho \in [0,1]} e^{- N (E_{0,N} (\rho, \mathbf{Q}_N, \mathbf{s})- \rho R)} .
\end{split}
\end{equation*}
The inequality in this expression comes from a direct application of Proposition~\ref{proposition:UpperBound}.
Following our previous observation that $E_{0,N} (\rho, \mathbf{Q}_N, \mathbf{s})$ only depends on $\mathbf{s}$ through its empirical distribution $\mathcal{T} (\mathbf{s})$, we can rewrite this upper bound as
\begin{equation*}
\begin{split}
P_{\mathrm{e}, S_N | S_0} (s_N | s_0)
&\leq \sum_{T \in \type}
\Pr \left( \type (\mathbf{S}) = T, S_N = s_N | S_0 = s_0 \right)
% SINGLE
\min_{\rho\in [0,1]} e^{- N (E_{0,N} (\rho, \mathbf{Q}_N, T) - \rho R)} .
\end{split}
\end{equation*}
The first bound in Proposition~\ref{theorem:EmpiricalDistributionBound} is equivalent to this inequality, yet it is expressed using a more concise notation.
The second bound holds because the sum of non-negative minimums is upper bounded by the minimum of the summands.
\end{IEEEproof}

In words, this result is obtained by first grouping channel state sequences according to types, applying an exponential upper bound on the probability of decoding failure to each group, and then taking an expectation over possible empirical distributions.
From a large deviations perspective, this decomposition into summands is pertinent because it can be employed to identify the dominating behavior of the system as block length becomes increasingly large.
Interestingly, the upper bound on the error probability only depends on the initial and final states of the channel through the empirical distributions.

Next, we consider a Gilbert-Elliott type channel, where the cardinality of the channel state space is $\mathcal{S}$.
Let sequence $\mathbf{s}$ be fixed and recall that $s_n \in \{ 1, 2, \ldots, \mathcal{S} \}$.
Then, by Proposition~\ref{proposition:UpperBound}, we get
\begin{equation} \label{equation:RelativeFrequenciesExpression}
\begin{split}
% SINGLE
E_{0,N} (\rho, \mathbf{Q}_N, \mathbf{s})
&= - \frac{1}{N} \ln \prod_{n=0}^{N-1} \frac{1}{2^{\rho}}
\left( \varepsilon_{s_n}^{\frac{1}{1+\rho}}
+ ( 1 - \varepsilon_{s_n} )^{\frac{1}{1+\rho}} \right)^{1+\rho} \\
&= - \frac{1}{N} \sum_{i=1}^{\mathcal{S}} n_i \ln \frac{1}{2^{\rho}}
\left( \varepsilon_i^{\frac{1}{1+\rho}}
+ (1 - \varepsilon_i )^{\frac{1}{1+\rho}} \right)^{1+\rho}
% SINGLE
= \sum_{i=1}^{\mathcal{S}}\frac{n_i}{N} b_i (\rho) ,
\end{split}
\end{equation}
where $n_i$ is the number of visits to channel state~$i$ in sequence $\mathbf{s}$, $\nicefrac{n_i}{N}$ is the fraction of time spent in state~$i$, and
\begin{equation} \label{equation:ConstantBi}
b_i (\rho) = - \ln \frac{1}{2^{\rho}}
\left( \varepsilon_i^{\frac{1}{1+\rho}}
+ \left( 1 - \varepsilon_i \right)^{\frac{1}{1+\rho}} \right)^{1+\rho} .
\end{equation}
Without loss of generality, we assume the error probabilities in different states are ordered such that
\begin{equation*}
\epsilon_1 < \epsilon_2 <\cdots<\epsilon_{\mathcal{S}} \leq \frac{1}{2},
\end{equation*}
which implies
\begin{equation} \label{eq:channelordering}
b_1 (\rho) > b_2 (\rho) >\cdots> b_{\mathcal{S}}(\rho).
\end{equation}

A key observation from \eqref{equation:RelativeFrequenciesExpression} is that the upper bound in \eqref{eq:mainbound} can be rewritten as an expectation with respect to the distribution of a weighted sum of the state occupation times.
This observation is of significant importance, as it reduces the computational complexity of the bound in \eqref{eq:mainbound} by a great amount.
The following remark precisely characterizes the statement.

\begin{remark}
Suppose the random variable $N_i$ denotes the number of visits to channel state i over the duration of a codeword, and let
\begin{equation} \label{eq:Wrv}
%W(\rho)=E_{0,N} (\rho, \mathbf{Q}_N, \mathbf{s})=\sum_{i=1}^{\mathcal{S}}\eta_i b_i (\rho),
W(\rho)=\frac{1}{N}\sum_{i=1}^{\mathcal{S}}N_i b_i (\rho),
\end{equation}
designate the weighted sum of the normalized occupancy times.
Then, the second upper bound on the error probability in \eqref{eq:mainbound} can be rewritten as
\begin{align}
\min_{\rho \in [0,1]} \sum_w
P_{W(\rho), S_N | S_0} (w, s_N | s_0)
e^{-N (w - \rho R)},\label{eq:boundwtsum}
\end{align}
where $P_{W(\rho), S_N | S_0} (w, s_N | s_0)$ represents the probability that $W(\rho) = w$ and $S_N = s_N$, given initial state $S_0 = s_0$.
We emphasize that the weights $b_i (\rho)$ are functions of the optimizing parameter $\rho$ and are deterministic.
\end{remark}

We note that the exponential term in \eqref{eq:mainbound} is averaged with respect to the joint distribution of channel state occupations $P_{\type (\mathbf{S}), S_N | S_0} (T, s_N | s_0)$.
%However, given the distribution of $E_{0,N} (\rho, \mathbf{Q}_N, \mathbf{s})$, which is the weighted sum of the normalized occupation times, the bound is
However, conditioned on $W(\rho)$, the bound is statistically independent of $\type (\mathbf{S})$. %$P_{\type (\mathbf{S}), S_N | S_0} (T, s_N | s_0)$.
In fact, when the channel state information is available at the receiver, $W(\rho)$ is a sufficient statistic to compute the upper bound on probability of decoding error.
The joint distribution of the channel states provides more information than what is needed to derive the bound.
This key characterization significantly improves the computational efficiency of the bounding technique, especially when the number of channel states increases.

%For illustrative purposes, we compute the upper bounds introduced in \eqref{eq:mainbound} for the Gilbert-Elliott channel.
%The channel ordering $\epsilon_1 < \epsilon_2 \leq \frac{1}{2}$ implies $b_1 (\rho) \geq b_2 (\rho)$.
For illustrative purposes, we derive the upper bounds on $P_{\mathrm{e}, S_N | S_0} (s_N|s_0)$ for the two-state Gilbert-Elliott channel.
Exploiting the Markov structure of this channel, we get
\begin{equation} \label{equation:ErrorBound}
\begin{split}
% SINGLE
P_{\mathrm{e}, S_N | S_0} (s_N  | s_0)
&\leq \min_{\rho \in [0,1]}\sum_{\mathbf{s}}
e^{- N ( E_{0,N} (\rho, \mathbf{Q}_N, \mathbf{s}) - \rho R ) }
\Pr ( \mathbf{S} = \mathbf{s}, S_N = s_N | S_0 = s_0) \\
&= \min_{\rho \in [0,1]}\left( \mathbf{e}_{s_0} \begin{bmatrix} a(1,1) & a(1,2) \\
a(2,1) & a(2,2) \end{bmatrix}^{N} \mathbf{e}_{s_N}^{\mathrm{T}} \right)
e^{\rho N R } .
\end{split}
\end{equation}
In this equation, $\mathbf{e}_i$ represents the unit vector of length two with a one in the $i$th position.
Matrix entries are defined by $a(i,j) = \left[ \mathbf{P} \right]_{ij} e^{b_i}$, where the transition probability matrix $\mathbf{P}$ is given in \eqref{equation:ProbabilityTransitionMatrix}.
We emphasize that this inequality holds for any $\rho \in [0, 1]$ and, hence, the bound can be tightened by minimizing over $\rho$.

\begin{remark}
The bound in~\eqref{equation:ErrorBound} is very similar to Gallager's exponential bound for finite-state channels~\cite[Thm.~5.9.3, pp.~185]{Gallager-1968}  when the receiver has perfect state information.
The main difference is that Gallager considers the ergodic regime and his equation simplifies to the logarithm of the largest eigenvalue of the matrix.
We omit this simplification because we are mainly interested in non-asymptotic regimes.
\end{remark}

\section{The Rare-Transition Regime}
\label{section:RareTransitionRegime}

In a traditional setting where $\mathbf{P}$ is kept constant, the upper bound given in \eqref{eq:mainbound} can be refined using the Perron-Frobenius theorem~\cite[pp.~184--185]{Gallager-1968}.
%, as described in Section~5.9 of \cite[pp.~184--185]{Gallager-1968}.
The more intriguing scenario for our purpose is the rare-transition regime where individual transition probabilities vary with $N$.
We introduce such a dependency through the sampling of a continuous-time Markov chain (CTMC) $\mathcal{X}(\cdot)$, defined by the infinitesimal generator matrix $\mathbf{Q}$.

Following conventional notation, we use $\Omega$ to designate the sample space, and we represent a generic outcome by $\omega$.
Whenever needed, we use superscript $\omega$ to refer to a particular realization.
For instance, $\mathcal{X}(\cdot)$ denotes the CTMC whereas $\mathcal{X}^{\omega}(\cdot)$ symbolizes the sample path associated with realization $\omega$.
In particular, $\mathcal{X}^{\omega}(\cdot)$ defines a mapping from the time interval $[0,\infty)$ to the state space $\{1,2,\ldots,\mathcal{S}\}$.
The need for this notation will become manifest shortly.

Suppose $\mathcal{X}(\cdot)$ is sampled at every $\nicefrac{1}{N}$ unit of time.
Then, we can construct a continuous-time version of the sampled chain as follows,
\begin{equation} \label{equation:SampledVersion}
\mathcal{X}_N (t) = \mathcal{X} \left(\frac{\lfloor Nt \rfloor}{N} \right) .
\end{equation}
Let $\mathbf{P}_N$ represent the transition probability matrix of the sampled Markov chain given by $\mathcal{X}_{N}(\nicefrac{n}{N})$, $n\in \mathbb{N}$.
Matrix $\mathbf{P}_N$ is governed by $\mathbf{Q}$ through the equation~\cite[Thms.~2.1.1~\&~2.1.2]{Norris-98}
\begin{equation} \label{equation:RareProbabilityTransitionMatrix}
\mathbf{P}_N = \exp \left( \frac{\mathbf{Q}}{N} \right) .
\end{equation}
We note that $\mathbf{P}_N$ is also the transition probability matrix of the Markov chain $\mathcal{X}(t)$ for a time interval of length $\nicefrac{1}{N}$,
\begin{equation*}
\left[ \mathbf{P}_N \right]_{ij}
= \Pr \left( \mathcal{X} \left( \frac{1}{N} \right) = j | \mathcal{X}(0)=i \right) .
\end{equation*}
In a similar fashion, we can make a distinction between the sampled chain $\mathcal{X}_N$ and the realization $\mathcal{X}_N^{\omega}$ associated with outcome $\omega$.

As an example, consider a two-state Markov process.
In this case, the infinitesimal generator matrix can be written as
\begin{equation} \label{eq:Q}
\mathbf{Q}=\begin{bmatrix}
-\mu & \mu \\
\xi & - \xi
\end{bmatrix}
\qquad \mu, \xi >0,
\end{equation}
and, consequently, we can express the transition probability matrix of the sampled process as
\begin{equation} \label{eq:PN}
\mathbf{P}_N = \frac{1}{\mu+\xi}
\begin{bmatrix}
\xi + \mu e^{-\frac{\xi+\mu}{N}}
& \mu \left( 1-e^{-\frac{\xi+\mu}{N}} \right) \\
\xi \left( 1-e^{-\frac{\xi+\mu}{N}} \right)
& \mu + \xi e^{-\frac{\xi+\mu}{N}}
\end{bmatrix} .
\end{equation}
As seen above, jumps in the discrete chain become less likely when $N$ increases.
This should be expected because a refined sampling of the CTMC does not alter the character of the underlying process.
Furthermore, the roles of boundary states are preserved, a property which is key for our subsequent analysis.
At this point, it is instructive to note that the inequalities presented in Section~\ref{section:Modeling} apply in the context of rare transitions as well, albeit using $\mathbf{P}_N$ rather than a fixed $\mathbf{P}$.

An important benefit of the rare-transition regime is the existence of approximate error bounds that can be computed efficiently.
These approximate bounds can be obtained using the following steps.
First, we show that the distributions of the occupation times for the sampled Markov chains converge to the distribution of the channel state occupation times of the original CTMC, as the sampling interval $\nicefrac{1}{N}$ decreases to zero.
Second, we employ standard results pertaining to the convergence of empirical measures to get approximate upper bounds on the probabilities of decoding failure at the destination, using the continuous measures.
We then leverage a numerical procedure to compute the distributions of weighted sums of channel state occupations for the CTMC~\cite{Sericola-sm00,Bladt-mbm02}.
Collecting these results, we arrive at the desired characterization of channels with memory.

Recall the CTMC $\mathcal{X} (t)$ and its sampled variants $\mathcal{X}_N(t)$, as introduced in \eqref{equation:SampledVersion}.
For every channel state~$i$, we define the occupation times pathwise through the integrals below,
\begin{align*}
\eta_i^{\omega}
&= \int_0^1 \boldsymbol{1}_{ \left\{ \mathcal{X}^{\omega}(t)=i \right\} } dt \\
\eta_{N,i}^{\omega}
&= \int_0^1 \boldsymbol{1}_{ \left\{ \mathcal{X}_N^{\omega}(t)=i \right\} } dt
= \frac{1}{N} \sum_{k=1}^N \boldsymbol{1}_{ \left\{ \mathcal{X}_N^{\omega}({k}/{N})=i \right\} },
\end{align*}
where $\boldsymbol{1}_{\{\cdot\}}$ denotes the standard indicator function.
Having specified the values of the occupation times for every possible outcome $\omega$, these equations unambiguously define random variables $\eta_i$ and $\eta_{N, i}$.

\begin{proposition} \label{proposition:MarkovConvergence}
The sequence of random vectors given by
\begin{equation*}
\overline{\eta}_N
= \left( \eta_{N,1}, \ldots, \eta_{N,\mathcal{S}} \right) ,
\qquad N = 1, 2, \ldots
\end{equation*}
converges almost surely to random vector $\overline{\eta} = \left( \eta_1, \ldots, \eta_{\mathcal{S}} \right)$ as $N$ approaches infinity.
\end{proposition}
\begin{IEEEproof}
The CTMC $\mathcal{X}(t)$ is time-homogeneous and its state space has finite cardinality.
It is therefore non-explosive~\cite[Sec.~2.7]{Norris-98}.
This implies that the number of transitions in the interval $[0,1]$ is finite almost surely; we can then write $\Pr (\Omega') = 1$, where
\begin{equation*}
\Omega' = \{ \omega \in \Omega | \mathcal{X}^{\omega} (t)
\text{ has finitely many jumps in }[0,1] \} .
\end{equation*}
For any $\omega \in \Omega'$, the function $\mathcal{X}^{\omega}$ is bounded and continuous almost everywhere on $[0,1]$.
It therefore fulfills Lebesgue's criterion for Riemann integrability~\cite[pp. 323]{Rudin-76} and, as such,
\begin{equation*}
\begin{split}
\eta_i^{\omega}
&= \int_0^1 \boldsymbol{1}_{ \left\{\mathcal{X}^{\omega}(t)=i \right\} } dt \\
&= \lim_{N \rightarrow \infty}
\frac{1}{N} \sum_{k=1}^N \boldsymbol{1}_{ \left\{\mathcal{X}_N^{\omega}(k/N)=i \right\}}
= \lim_{N \rightarrow \infty} \eta_{N,i}^{\omega} .
\end{split}
\end{equation*}
Since the number of channel states is finite, this result readily extends to vectors,
\begin{equation*}
\lim_{N \rightarrow \infty}
d_1 \left( \overline{\eta}_N^{\omega}, \overline{\eta}^{\omega} \right)
= 0 \qquad \forall \omega \in \Omega',
\end{equation*}
where $d_1(\cdot, \cdot)$ is the $\ell_1$~distance on $\mathbb{R}^{\mathcal{S}}$.
Equivalently, we can write
\begin{equation*}
\Pr \left( \omega \in \Omega \middle| \lim_{N \rightarrow \infty} d_1 \big( \overline{\eta}_N^{\omega}, \overline{\eta}^{\omega} \big) = 0 \right) = 1 .
\end{equation*}
That is, $\overline{\eta}_N$ converges to $\overline{\eta}$ almost surely, as desired.
\end{IEEEproof}

Almost sure convergence implies convergence in probability and in distribution~\cite[Sec.~8.5]{Resnick-99}.
Thus, from Proposition~\ref{proposition:MarkovConvergence}, we gather that $\overline{\eta}_N$ converges to $\overline{\eta}$ in distribution, which is sufficient for our later purpose.
This further implies that the occupation times of a sequence of independent discrete-time Markov chains, each generated according to $\mathbf{P}_N$, converge in distribution to $\overline{\eta}$.
That is, the sampling of the CTMC described above is a powerful mathematical tool to get the result we seek.

In the discussion above, we are intentionally vague about the probability laws on $\Omega$.
The careful reader will note that our arguments apply to the equilibrium distribution of the Markov chain as well as the conditional measure where the Markov process starts in state~$i \in \{ 1, 2, \ldots, \mathcal{S} \}$, at time zero.
Moreover, by extension, these findings apply to probabilities where the final channel state is taken into account.
To distinguish between these different scenarios, we introduce a shorthand notation for these joint probabilities,
\begin{gather*}
\begin{split}
F  (r_1, r_2, \ldots, r_{\mathcal{S}-1}) 
&= \Pr \left( \eta_1 \leq r_1, \ldots, \eta_{\mathcal{S}-1} \leq r_{\mathcal{S}-1} \right)
\end{split} \\
% SINGLE
\begin{split}
F_{ij} (r_1, r_2, \ldots, r_{\mathcal{S}-1})
&= \Pr \left( \eta_1 \leq r_1, \ldots, \eta_{\mathcal{S}-1} \leq r_{\mathcal{S}-1}, S_{\mathrm{f}} = j , S_{\mathrm{i}} = i \right)
\end{split} \\
% SINGLE
\begin{split}
F_{j|i} (r_1, r_2, \ldots, r_{\mathcal{S}-1})
&= \Pr \left( \eta_1 \leq r_1, \ldots, \eta_{\mathcal{S}-1} \leq r_{\mathcal{S}-1}, S_{\mathrm{f}} = j | S_{\mathrm{i}} = i \right) .
\end{split}
\end{gather*}
In our labeling, $S_{\mathrm{i}}$ identifies the initial state of the channel and $S_{\mathrm{f}}$ specifies its final value.
We can define $F_N$, $F_{N,ij}$ and $F_{N, j|i}$ in an analogous manner.
It is immediate from Proposition~\ref{proposition:MarkovConvergence} that $dF_N \Rightarrow dF$, $dF_{N,ij} \Rightarrow dF_{ij}$, and $dF_{N, j|i} \Rightarrow dF_{j|i}$ as $N$ grows to infinity, where the symbol $\Rightarrow$ denotes convergence in distribution.
%Key to our upcoming analysis is the fact that the distribution of $\overline{\eta}_N$ is straightforward to compute using $\mathbf{P}_N$, whereas the exact distribution of $\overline{\eta}$ is much more difficult to acquire.
As a direct consequence of Proposition~\ref{proposition:MarkovConvergence}, we can apply these results to affine combinations of $\eta_1, \ldots, \eta_{\mathcal{S}}$.

\begin{corollary}\label{cor_W}
Let $\rho$ be fixed and recall the coefficients $b_i (\rho)$ found in \eqref{equation:ConstantBi}.
The sequence of random variables given by
\begin{equation*}
W_N (\rho) = \sum_{i=1}^{\mathcal{S}} \eta_{N,i} b_i (\rho)
\end{equation*}
converges in distribution to random variable
\begin{equation*}
W(\rho) = \sum_{i=1}^{\mathcal{S}} \eta_i b_i (\rho)
\end{equation*}
as $N$ approaches infinity.
\end{corollary}

We point out that the expression for $W(\rho)$ above differs slightly from \eqref{eq:Wrv}, as it reflects the notation developed for the current asymptotic setting.
Again, this result is valid for the probability laws associated with $F$, $F_{ij}$ and $F_{j|i}$.
The weighted sum $W (\rho)$ is of such importance in our impending discussion that we introduce a convenient notation for its corresponding probability laws as well,
\begin{align*}
G (w) &= \Pr (W(\rho) \leq w) \\
G_{ij} (w) &= \Pr (W(\rho) \leq w , S_{\mathrm{i}} = i, S_{\mathrm{f}} = j) \\
G_{j|i} (w) &= \Pr (W(\rho) \leq w , S_{\mathrm{f}} = j | S_{\mathrm{i}} = i) .
\end{align*}
In addition, we write $G_N$, $G_{N, ij}$ and $G_{N, j|i}$ for the measures associated with $W_N (\rho)$.
It may be helpful to point out that we introduce a slight abuse of notation in establishing these quantities; the dependence of these probability laws on $\rho$ is implicit.
This is intentional as the alternative makes the notation overly cumbersome and confusing.
In the limiting case, the measures $G$, $G_{ij}$, and $G_{j|i}$ are continuous almost everywhere.
Wherever needed, we can emphasize the dependence on $\rho$ by writing
\begin{align*}
dG (w) &= f_{W(\rho)} (w) dw \\
dG_{ij} (w) &= f_{W(\rho), S_{\mathrm{i}}, S_{\mathrm{f}}} (w, i, j) dw \\
dG_{j|i} (w) &= f_{W(\rho), S_{\mathrm{f}} | S_{\mathrm{i}}} (w, j | i) dw
\end{align*}
where $f(\cdot)$ is our generic representation of a probability density, possibly with weighted Dirac delta components.

\begin{proposition} \label{proposition:ApproximateUpperBound}
Suppose that a message is transmitted over a fading channel with $\mathcal{S}$ states, using the random coding scheme introduced in Section~\ref{section:Modeling}.
An approximate upper bound for the error probability $P_{\mathrm{e}, S_N | S_0} (j | i)$ is given by
\begin{equation} \label{equation:ApproximateUpperBound}
\begin{split}
P_{\mathrm{e}, S_N | S_0} (j | i)
\lesssim \int_{b_{\mathcal{S}}(\rho)}^{b_1(\rho)} \! & \min \left\{ 1, e^{ - N ( w - \rho R ) } \right\}
f_{W(\rho), S_{\mathrm{f}} | S_{\mathrm{i}}} (w, j | i) dw .
% SINGLE
\end{split}
\end{equation}
The approximation in \eqref{equation:ApproximateUpperBound} reflects its potential use in selecting efficient coding schemes.
The precise mathematical meaning underlying this equation is described in the proof.
\end{proposition}
\begin{IEEEproof}
\begin{comment}
In order to proof Proposition~\ref{proposition:ApproximateUpperBound}, we first prove the following bound
\begin{equation} \label{equation:ApproximateUpperBoundJoint}
\begin{split}
P_{\mathrm{e}, S_N , S_0} (j , i)
\lesssim \int_{b_{\mathcal{S}}(\rho)}^{b_1(\rho)} \! & \min \left\{ 1, e^{ - N ( W(\rho) - \rho R ) } \right\} \\
&\times f_{W(\rho), S_{\mathrm{f}} , S_{\mathrm{i}}} (w, j , i) dw
\end{split}.
\end{equation}
\end{comment}
Let $\rho \in (0,1)$ be fixed.
We know from Corollary~\ref{corollary:UpperBound} that, for channel type~$T$, the error probability is bounded by
\begin{equation*}
P_{\mathrm{e} | \mathbf{S}} (T)
\leq e^{- N \left( E_{0, N} (\rho, \mathbf{Q}_N, T) - \rho R \right) }
%&= e^{- N \left( W(\rho) - \rho R \right)} .
= e^{- N \left( w - \rho R \right)} .
\end{equation*}
where $w = \sum_{i=1}^{\mathcal{S}} \frac{n_i}{N} b_i (\rho)$ is determined by the channel type.
%We emphasize that the second expression is the error exponent specialized to the two-state channel, as derived in \eqref{equation:RelativeFrequenciesExpression}.
We can readily tighten this bound to
\begin{equation*}
%P_{\mathrm{e} | \mathbf{S}} (T) \leq \min \left\{ 1, e^{- N \left( W(\rho) - \rho R \right)}\right\}
P_{\mathrm{e} | \mathbf{S}} (T) \leq \min \left\{ 1, e^{- N \left( w - \rho R \right)}\right\}
\end{equation*}
because individual probabilities cannot exceed one.
It is useful to point out that the expression $w - \rho R$ is an affine, strictly increasing function of $w$.
%Moreover, based on our channel state ordering \eqref{eq:channelordering}, the above function is strictly increasing with respect to $r_1$.
For the purpose of exposition, let $g_N(w)$ be defined by %(r_1,r_2,\ldots,r_{\mathfrak{S}-1})
\begin{equation*}
\begin{split}
g_N (w) &=  \min \left\{ 1,
e^{- N \left( w - \rho R \right)} \right\}
% SINGLE
= \begin{cases} e^{- N \left( w - \rho R \right)} & w < \varrho \\
1 & w \geq \varrho \end{cases} ,
\end{split}
\end{equation*}
where the threshold $\varrho=\rho R$.
The sequence of functions $\{ g_N (w) \}$ converges pointwise to
\begin{equation*}
g(w) = \mathbf{1}_{[\varrho, \infty)} (w)
= \begin{cases} 0 & w < \varrho \\
1 & w \geq \varrho \end{cases} ,
\end{equation*}
which is uniformly continuous on the set $w \in [0,b_1(\rho)] \setminus \{\varrho\}$.
By taking an expectation over $W(\rho)$, we get
%the distribution of $T$ for initial state~$i$ and terminating state~$j$, we get
\begin{equation} \label{equation:ExpectationUpperBound}
\begin{split}
% SINGLE
P_{\mathrm{e}, S_N | S_0} (j | i)
%&\leq \int_{b_{\mathcal{S}}(\rho)}^{b_1(\rho)} \min \left\{ 1, e^{- N \left( \sum_{k=1}^{\mathcal{S}}\frac{n_k}{N} b_k(\rho)- \rho R \right)} \right\} dG_{N,j|i} \\
&\leq \int_{b_{\mathcal{S}}(\rho)}^{b_1(\rho)} \min \left\{ 1, e^{- N (w - \rho R)} \right\} dG_{N,j|i}(w) \\
&= \sum_{w} P_{W_N(\rho), S_N | S_0} (w, j | i) \min \left\{ 1, e^{-N (w - \rho R)} \right\} .
\end{split}
\end{equation}
While this upper bound can be computed numerically if the distribution of $W_N(\rho)$ is known, we are also interested in approximations that provide good intuition as well as computational efficiency at the expense of a little accuracy.

As a next step, we will establish that
\begin{equation} \label{equation:LimitingIntegral}
\lim_{N \rightarrow \infty} \int_{b_{\mathcal{S}}(\rho)}^{b_1(\rho)} g_N \, dG_{N,j|i}
= \int_{b_{\mathcal{S}}(\rho)}^{b_1(\rho)} g \, dG_{j|i} .
\end{equation}
\iftrue
From Corollary~\ref{cor_W}, we know that the set of measures $\{ G_{N,j|i} \}$ converges in distribution to $G_{j|i}$.
Also, for any converging sequence $w_N \in [b_{\mathcal{S}}(\rho), b_1(\rho)]$ with $\lim_{N \rightarrow \infty} w_N = w \neq \varrho$, we have $g_N (w_N) \rightarrow g(w)$.
This is pertinent because the limiting measure $G_{j|i} (w)$ is continuous at $\varrho \in (b_{\mathcal{S}}(\rho),b_1(\rho))$ and, consequently, the event $\{ w = \varrho \}$ has probability zero.
Collecting these observations, we can can apply~\cite[Thm.~5.5]{Billingsley-1968} and thereby establish the validity of \eqref{equation:LimitingIntegral}.
\else
To do so, we consider the expression
\begin{equation} \label{equation:IntegralDifference}
\begin{split}
&\int g_N \, dG_{ij}^{(N)} - \int g \, dG_{ij} \\
&=\int (g_N - g) \, dG_{ij}^{(N)}
+ \int g \, dG_{ij}^{(N)}
- \int g \, dG_{ij} \\
\end{split}
\end{equation}
The first integral in the last summation can be written as
\begin{equation*}
\begin{split}
&\int_0^1 (g_N - g) \, dF_{ij}^{(N)}
%= \int_0^{\tau}  g_N \, dF_{ij}^{(N)}
= \int_0^{\tau - \delta}  g_N \, dF_{ij}^{(N)}
+ \int_{\tau - \delta}^{\tau}  g_N \, dF_{ij}^{(N)}
\end{split}
\end{equation*}
where $\delta > 0$.
We notice that
\begin{equation*}
\int_0^{\tau - \delta}  g_N \, dF_{ij}^{(N)}
\leq g_N (\tau - \delta) \int_0^{\tau - \delta} \, dF_{ij}^{(N)}
\rightarrow 0
\end{equation*}
as $N \rightarrow \infty$.
Moreover, since $F_{ij}^{(N)}$ converges weakly to $F_{ij}$, we have
\begin{equation*}
\begin{split}
\int_{\tau - \delta}^{\tau}  g_N \, dF_{ij}^{(N)}
\leq F_{ij}^{(N)} ((\tau - \delta, \tau))
\rightarrow F_{ij} ((\tau - \delta, \tau))
\end{split}
\end{equation*}
as $N \rightarrow \infty$.
This follows from a straightforward application of the Portmanteau Theorem~\cite[Thm.~8.4.1]{Resnick-99}.
The contribution of $F_{ij} ((\tau - \delta, \tau))$ can be made arbitrarily small by selecting a positive value for $\delta$ appropriately.
Collecting these results, we deduce that
\begin{equation*}
\lim_{N \rightarrow \infty} \int_0^1 (g_N - g) \, dF_{ij}^{(N)} = 0.
\end{equation*}
The second and third integrals in \eqref{equation:IntegralDifference} can be expressed as
\begin{equation*}
\begin{split}
\int_0^1 g \, dF_{ij}^{(N)}
- \int_0^1 g \, dF_{ij}
= F_{ij}^{(N)} ([\tau, 1]) - F_{ij} ([\tau, 1]) ,
\end{split}
\end{equation*}
yet this difference also vanishes as $N$ increases to infinity.
Putting these pieces together, we gather that \eqref{equation:LimitingIntegral} holds, as desired.
\fi

In view of these results, we can write
\begin{equation}
\int_{b_{\mathcal{S}}(\rho)}^{b_1(\rho)} g_N \, dG_{N,j|i}
\approx \int_{b_{\mathcal{S}}(\rho)}^{b_1(\rho)} g_N \, dG_{j|i}
\approx \int_{b_{\mathcal{S}}(\rho)}^{b_1(\rho)} g \, dG_{j|i}\label{eq:integrals}
\end{equation}
for large enough values of $N$.
\begin{comment}
In fact, since $g_N(w)$ is bounded and continuous,
\begin{equation*}
\begin{split}
&\left| \int_{b_{\mathcal{S}}(\rho)}^{b_1(\rho)} g_N dG_{N,j|i}
- \int_{b_{\mathcal{S}}(\rho)}^{b_1(\rho)}g_N dG_{j|i} \right| \\
&\leq \sup_w | g_N(w) |
\left| \int_{b_{\mathcal{S}}(\rho)}^{b_1(\rho)} dG_{N,j|i}
- \int_{b_{\mathcal{S}}(\rho)}^{b_1(\rho)} dG_{j|i} \right| \\
&\leq \left| \int_{b_{\mathcal{S}}(\rho)}^{b_1(\rho)} dG_{N,j|i}
- \int_{b_{\mathcal{S}}(\rho)}^{b_1(\rho)} dG_{j|i} \right| .
\end{split}
\end{equation*}
Since $\{ G_{N,j|i} \}$ converges in distribution to $G_{j|i}$, it follows that the last upper bound above vanishes as $N$ increases to infinity.
\end{comment}
\begin{comment}
\begin{align*}
\left|\int_{b_{\mathcal{S}}(\rho)}^{b_1(\rho)} g_N \, dG_{N,j|i}-\int_{b_{\mathcal{S}}(\rho)}^{b_1(\rho)}g_N \, dG_{j|i}\right|&\le\\
\left(\max_w g_N(w)\right)&d_{TV}(G_{N,j|i},G_{j|i})\\
&\le d_{TV}(G_{ij}^{(N)},G_{ij}),
\end{align*}
and $d_{TV}(G_{ij}^{(N)},G_{ij})\rightarrow 0$.
\end{comment}
Since the first integral in \eqref{eq:integrals} provides an upper bound on $P_{\mathrm{e}, S_N | S_0} (j | i)$, the approximate upper bound of \eqref{equation:ApproximateUpperBound} immediately follows.
That is, if the code length is large enough, then the approximation in Proposition~\ref{proposition:ApproximateUpperBound} is justified.
\end{IEEEproof}

Using \eqref{equation:ApproximateUpperBound} to select system parameters gives an alternate, computationally efficient way to design good systems.
We emphasize that one does not necessarily need to compute a sequence of distributions for $\{ W_N (\rho) \}$ to follow this solution path.
Rather, it is possible to accurately approximate the distribution of $W(\rho)$ directly using an iterative approach.
We review one such method below; it leverages numerical techniques introduced in \cite{Sericola-sm00,Bladt-mbm02} to compute the distributions of reward functions on CTMCs.
This method applies to channels with arbitrary, yet finite numbers of states.
In contrast, the standard approach associated with \eqref{equation:ExpectationUpperBound} entails computing $G_{N,j|i}$ explicitly for multiple values of $N$, a cumbersome task.

Proposition~\ref{proposition:NumericalComputationW} presents a numerical method to compute the distribution of $W(\rho)$.
This method is adapted from~\cite{Sericola-sm00,Bladt-mbm02} and, as such, it is presented without detailed proof.
In practice, the infinite sum needs to be truncated according to an appropriate criterion.
To present this result, we need to introduce relevant notations.
Let $\mathcal{A}$ be the transition matrix given by
\begin{equation*}
\mathcal{A} = \mathbf{I} + \frac{\mathbf{Q}}{\sigma}
\end{equation*}
where $\mathbf{I}$ is the identity matrix, $\sigma \geq \max_{k} | \mathbf{Q}_{kk}|$, $k \in \{ 1, 2, \ldots, \mathcal{S} \}$, is a constant and $\{ \mathbf{Q}_{kk} \}$ are the diagonal elements of $\mathbf{Q}$.
Also, define matrix $\mathbf{G} (w)$ by
\begin{equation*}
\left[ \mathbf{G} (w) \right]_{ij} = G_{j|i} (w)
\end{equation*}
where $i, j \in \{ 1, 2, \ldots, \mathcal{S} \}$.

\begin{proposition} \label{proposition:NumericalComputationW}
Let $\rho$ be fixed and suppose that the channel is initially in state $S_{\mathrm{i}} = i$.
The probabilities of the events $\{ W (\rho) \leq w , S_{\mathrm{f}} = j | S_{\mathrm{i}} = i\}$ as functions of $w$ are continuous almost everywhere, and they have at most $\mathcal{S}$ discontinuities, with possible locations $b_{\mathcal{S}} (\rho), \ldots, b_1(\rho)$.
Furthermore, for $w \in [b_k (\rho), b_{k-1} (\rho))$ and $2 \leq k \leq \mathcal{S}$, we have
\begin{equation*}
\mathbf{G} (w)
= \sum_{n=0}^\infty e^{-\sigma} \frac{\sigma^n}{n!}
\sum_{l=0}^n \binom{n}{l} w_k^l(1 - w_k)^{n-l} \mathbf{C}^{(k)}(n,l),
\end{equation*}
where
\begin{equation*}
w_k = \frac{w - b_k (\rho)}{b_{k-1} (\rho) - b_k (\rho)} .
\end{equation*}
The matrices $\left\{ \mathbf{C}^{(k)} (n, l) \right\}$ are defined component-wise by
\begin{equation*}
\left[ \mathbf{C}^{(k)} (n,l) \right]_{cd} = C^{(k)}_{cd} (n,l)
\qquad c,d \in \{ 1,2,\ldots,\mathcal{S} \} ,
\end{equation*}
and the individual entries in each of these matrices are given by the following two recurrence relations.

For $k \leq c \leq \mathcal{S}$ and $0 \leq d \leq \mathcal{S}$,
\begin{equation*}
\begin{split}
C^{(k)}_{cd} (n,l)
&= \frac{b_k (\rho) - b_c (\rho)}{b_{k-1} (\rho) - b_c (\rho)} C^{(k)}_{cd} (n,l-1) \\
&+ \frac{b_{k-1} (\rho) - b_k (\rho)}{b_{k-1} (\rho) - b_c (\rho)} \sum_{e=0}^{\mathcal{S}} \left[\mathcal{A}\right]_{ce} C^{(k)}_{ed} (n-1,l-1)
\end{split}
\end{equation*}
where $1 \leq l \leq n$. 
For $n \geq 0$ and $k > 1$, we apply the boundary conditions $C^{(1)}_{cd} (n,0)=0$ and $C^{(k)}_{cd} (n,0) = C^{(k-1)}_{cd} (n,n)$.

Similarly, for $0 \leq c \leq k-1$ and $0 \leq d \leq \mathcal{S}$,
\begin{equation*}
\begin{split}
C^{(k)}_{cd} (n,l)
&= \frac{b_c (\rho) - b_{k-1} (\rho)}{b_c (\rho) - b_k (\rho)} C^{(k)}_{cd} (n,l+1)\\
&+\frac{b_{k-1} (\rho) - b_k (\rho)}{b_k (\rho) - b_c (\rho)}
\sum_{e=0}^{\mathcal{S}} \left[\mathcal{A}\right]_{ce} C^{(k)}_{ed} (n-1,l)
\end{split}
\end{equation*}
where $0 \leq l \leq n-1$.
In this case, for $n \geq 0$ and $k < \mathcal{S}$, we can write the boundary conditions $C^{(\mathcal{S})}_{cd} (n,n) = \left[ \mathcal{A}^n \right]_{cd}$ and $C^{(k)}_{cd} (n,n) = C^{(k+1)}_{cd}(n,0)$.
\end{proposition}
\begin{IEEEproof}[Sketch of proof]
We emphasize, again, that this proposition is adapted from a general technique found in~\cite{Sericola-sm00,Bladt-mbm02}.
In paralleling the argument presented therein, the weights $b_1 (\rho), \ldots, b_{\mathcal{S}} (\rho)$ play the role of reward rates and $W(\rho)$ represents the total continuous reward over the interval $[0, 1)$.
The possible discontinuities in $G_{j|i} (w)$ have to do with the non-vanishing probabilities that the chain does not visit certain states during time interval $[0, 1)$.
\end{IEEEproof}
\begin{comment}
The following corollary is a consequence of Proposition~\ref{proposition:NumericalComputationW}, and it specifies density functions for the smooth parts of matrix $\mathbf{G}$.

\begin{corollary}
For $w \in (b_{k}(\rho),b_{k-1}(\rho))$ and $1 \leq k \leq \mathcal{S}$, we can write
\begin{equation*}
\begin{split}
d\mathbf{G} (w)
= & \frac{\sigma e^{-\sigma}}{b_{k-1}(\rho) - b_{k}(\rho)}
\sum_{n=0}^\infty \frac{\sigma^n}{n!} \sum_{l=0}^n \binom{n}{l} w_k^l(1 - w_k)^{n-l} \\
&\times \left[ \mathbf{C}^{(k)} (n+1,l+1) - \mathbf{C}^{(k)}(n+1,l) \right] .
\end{split}
\end{equation*}
\end{corollary}
\end{comment}

The discrete-time Markov chain whose probability transition matrix is given by $\mathcal{A}$ is called a uniformized chain \cite{Ross-83,Pierre-99}.
This chain can be paired to a Poisson process with rate $\sigma$ to form a continuous-time Markov chain.
The resulting chain is stochastically equivalent to $\mathcal{X}(t)$ and, as such, it possesses the same invariant probability distribution~\cite{Sericola-sm00}.
Furthermore, the matrix $\mathbf{P}_N$ can be written as
\begin{equation*}
\mathbf{P}_N = e^{-\frac{\sigma}{N}}
\exp \left( \frac{\sigma \mathcal{A}}{N} \right) ,
\end{equation*}
or, alternatively,
\begin{equation*}
\left[ \mathbf{P}_N \right]_{ij}
= e^{-\frac{\sigma}{N}} \sum_{k=0}^\infty \frac{1}{k!}
\left(\frac{\sigma}{N}\right)^k
\left[\mathcal{A}^k\right]_{ij} .
\end{equation*}
While the uniformized chain and the sampled chain are both derived from $\mathbf{Q}$, there remains an important difference.
By construction, the uniformized chain never misses a transition in its corresponding continuous-time Markov chain; whereas the sampled chain with its periodic structure can overlook jumps associated with fast transitions.
This makes the uniformized chain a more suitable object in describing Proposition~\ref{proposition:NumericalComputationW}.

As a special case of Proposition~\ref{proposition:NumericalComputationW}, we turn to the situation where $\mathcal{S} = 2$.
Not surprisingly, occupation times for this simple scenario have been studied in the past, and explicit expressions for their distributions exist \cite{Gabriel-Biometrika59,Pedler-jap71,Kovchegov-spl10}.
Still, the distributions provided therein only account for an initial state, and they do not specify a final state.
We must therefore modify these results slightly to match the needs of our current framework.

\begin{lemma} \label{lemma:WeakConvergence}
Consider a continuous-time Markov chain whose generator matrix is given by \eqref{eq:Q}.
The joint distributions governing occupation times and the final state, conditioned on the initial state, can be written as follows
\begin{align*}
% SINGLE
f_{\eta_1, S_{\mathrm{f}} | S_{\mathrm{i}}} (r,1|1) &= e^{-\mu r - \xi (1-r)}
\left( \delta(1-r) + \sqrt{\frac{\mu \xi r}{1-r}}
I_1 \left( 2 \sqrt{\mu \xi r(1-r)} \right) \right) \\
f_{\eta_1, S_{\mathrm{f}} | S_{\mathrm{i}}} (r, 2 | 1) &= \mu e^{-\mu r - \xi (1-r)}
I_0 \left( 2 \sqrt{\mu \xi r(1-r)} \right) \\
f_{\eta_1, S_{\mathrm{f}} | S_{\mathrm{i}}} (r, 1 | 2) &= \xi e^{-\mu r - \xi (1-r)}
I_0 \left( 2 \sqrt{\mu \xi r(1-r)} \right) \\
f_{\eta_1, S_{\mathrm{f}} | S_{\mathrm{i}}} (r,2|2) &= e^{-\mu r - \xi(1-r)}
\left( \delta(r) + \sqrt{\frac{\mu \xi (1-r)}{r}}
I_1 \left( 2 \sqrt{\mu \xi r(1-r)} \right) \right)
\end{align*}
where $I_0 (\cdot)$ and $I_1 (\cdot)$ represent modified Bessel functions of the first kind defined in \cite[Lem.~2, pp.~386]{Pedler-jap71}.
\end{lemma}
\begin{IEEEproof}
See the appendix.
\end{IEEEproof}

\begin{comment}
The sequence of conditional distributions induced by the two-state Markov chains $F^{(N)}_{ij}$ converges weakly to positive measure $F_{ij}$.
\end{comment}
\begin{comment}
Consider a collection of two-state Markov chains where the transition probabilities of the $N$th system are governed by $\mathbf{P}_N$, as described in \eqref{equation:RareProbabilityTransitionMatrix}.
The joint probability associated with $\eta \leq r$ and the $i\!\to\!j$ state transition for the length-$N$ system is
\begin{equation*}
F^{(N)}_{ij} (r) = \Pr \left( \eta \leq r, S_N = j | S_0 = i \right) .
\end{equation*}
Moreover, consider the following densities
\end{comment}

The corresponding expressions for the sampled chain are presented in the following Lemma.

\begin{lemma} \label{thm:1}
Consider a two-state channel whose transition probability matrix is given by \eqref{equation:ProbabilityTransitionMatrix}.
Assume that the number of visits to each state is recorded for a period spanning $N$ consecutive channel realizations.
The joint distributions governing the channel type and its final state, conditioned on the initial state, can be written in terms of the Gaussian hypergeometric function ${}_2F_1 (\cdot, \cdot; \cdot; \cdot)$ defined in \cite[Lem.~1, pp.~383]{Pedler-jap71}.
In particular, for $m=1, 2, \ldots, N-1$, we get
\begin{align*}
% SINGLE
P_{N_1, S_N | S_0} (m, 1 | 1) &= (1-\alpha)^m (1-\beta)^{N-m}
\Big( {}_2F_1 (-N + m, -m; 1; \lambda)
- {}_2F_1 (-N + m + 1, -m; 1; \lambda) \Big) \\
P_{N_1, S_N | S_0} (m, 2 | 1) &= \frac{(1-\alpha)^{m-1} (1-\beta)^{N-m+1}\alpha}{(1-\beta)}
{}_2F_1 (-N + m, -m+1; 1; \lambda) \\
P_{N_1, S_N | S_0} (m, 1 | 2) &= \frac{(1-\alpha)^{m+1} (1-\beta)^{N-m-1}\beta}{(1-\alpha)}
{}_2F_1 (-N + m + 1, -m; 1; \lambda) \\
P_{N_1, S_N | S_0} (m, 2 | 2) &= (1-\alpha)^{m}(1-\beta)^{N-m}
\Big( {}_2F_1 (-N + m, -m; 1; \lambda)
- {}_2F_1 (-N + m, -m + 1; 1; \lambda) \Big)
\end{align*}
where $N_1$ is the number of visits to the first state, and $\lambda=\frac{\alpha \beta}{(1-\alpha)(1-\beta)}$.
Special consideration must be given to extremal cases.
In particular, we have
\begin{align*}
P_{N_1, S_N | S_0} (0, \cdot | 1) &= P_{N_1, S_N | S_0} (N, \cdot | 2) = 0 \\
P_{N_1, S_N | S_0} (0, 2 | 2) &= (1 - \beta)^N \\
P_{N_1, S_N | S_0} (N, 1 | 1) &= (1 - \alpha)^N .
\end{align*}
\end{lemma}
\begin{IEEEproof}
See the appendix.
\end{IEEEproof}

We can relate these two results through Proposition~\ref{proposition:MarkovConvergence} and the definition of $\mathbf{P}_N$ in \eqref{eq:PN}.
Let $\alpha$ and $\beta$ be the constants defined in Lemma~\ref{thm:1}, and consider the assignment
\begin{align*}
\alpha &= \frac{\mu}{\mu+\xi} \left( 1-e^{-\frac{\xi+\mu}{N}} \right) \\
\beta &= \frac{\xi}{\mu+\xi} \left( 1-e^{-\frac{\xi+\mu}{N}} \right) .
\end{align*}
Then, the discrete distributions specified above converge to the occupation times described in Lemma~\ref{lemma:WeakConvergence}, under the asymptotic scaling $\frac{N_1}{N} \rightarrow \eta$, as $N$ grow unbounded.
As a side observation, we emphasize that two of the limiting measures in Lemma~\ref{lemma:WeakConvergence} are not absolutely continuous with respect to the Lebesgue measure.
Still, these distributions are well-defined positive measures~\cite[Chap.~8]{Resnick-99}.
The discontinuities can be explained by the fact that, in the rare-transition limit, there is a positive probability of staying in the initial channel state for the whole block.

Using the limiting distributions in Lemma~\ref{lemma:WeakConvergence}, one can compute the approximate upper bound found in \eqref{equation:ApproximateUpperBound} for the two-state case.
We stress that, for this simple channel, $W(\rho) = (b_1(\rho)-b_2(\rho)) \eta + b_2(\rho)$ is an affine function of $\eta$.
Hence, the distribution of $W(\rho)$ can be derived in terms of $\eta$.
For this simple case, it is straightforward to compute the approximate bound using the empirical distribution of the state occupancies,
\begin{equation} \label{equation:ApproximateUpperBound2state}
\begin{split}
% SINGLE
P_{\mathrm{e}, S_N | S_0} (j | i)
\lesssim \int_0^1 & \min \left\{ 1, e^{ - N ( (b_1(\rho)-b_2(\rho)) r + b_2(\rho) - \rho R ) } \right\}
f_{\eta, S_{\mathrm{f}} | S_{\mathrm{i}}} (r, j | i) dr .
\end{split}
\end{equation}
Yet, as the number of states increases, dealing with the joint distribution of the state occupation times and integrating over multiple variables is an increasingly complex task.
This difficulty is bypassed when we use the distribution of the weighted sum of occupation times since, in this latter case, we are dealing with a single random variable as opposed to a random vector.

In general, getting the distributions of the occupation times for the discrete chains is not needed to apply the result of Proposition~\ref{proposition:ApproximateUpperBound}.
However, for the two-state channel, the distributions are available for both the continuous-time and the sampled chains.
It is then instructive to compute the exact upper bound in \eqref{equation:ExpectationUpperBound} using the distributions given by Lemma~\ref{thm:1}, and compare it to the approximate bound presented in \eqref{equation:ApproximateUpperBound2state}.
Numerical results for this comparison are presented in Section~\ref{sec:numerical}, thereby offering supporting evidence for our proposed methodology.
We note that, even for the simple two-state case, computing the approximate upper bound in \eqref{equation:ApproximateUpperBound2state} is considerably more efficient than calculating \eqref{equation:ExpectationUpperBound}.
As we will see in Section~\ref{sec:numerical}, the price to pay for this computational efficiency is a small loss in accuracy.

From an engineering point of view, we are interested in cases where $N$ is dictated by the code length of a practical coding scheme.
The approximate upper bound can be used to perform a quick survey of good parameters.
Then, the exact expression based on the hypergeometric function can be employed for fine tuning locally.
As a final note on this topic, we emphasize that these upper bounds can be tightened by optimizing over $\rho \in [0, 1]$.
This task entails repeated computations of the bounds, which partly explains our concerns with computational efficiency.

A significant benefit in dealing with the Gilbert-Elliott channel model is its tractability.
The remaining of this article is devoted to the analysis of error probability and the queueing behavior of systems built around this two-state channel.
The next sections are dedicated to exact derivation of probabilities of detected and undetected decoding failure.
This is an intermediate step to characterize system performance, and it allows for fair evaluation of the proposed bounding technique.
Furthermore, we also show how to bound the probability of undetected errors with slight modifications to the approximated upper bounds.
In Sections~\ref{sec:queue} and \ref{sec:Stochastic-Dominance}, we explore the queueing performance of the system, using exact expressions and upper bounds on the detected and undetected probabilities of decoding failure.

\section{Exact Probability of Decoding Failures}

It is possible to compute exact probabilities of decoding failure under various decision schemes for the two-state Gilbert-Elliott channel, due to its simplicity.
Consequently, in this case, we can assess how close the bounds and the true probabilities of error are from one another.
The hope is that, if the Gilbert-Elliott bounds are reasonably tight, then the upper bounds for general finite-state channels will also be good.
Of course, it may be impractical to compute exact probabilities of error for more elaborate channels.
Even for Gilbert-Elliot type channels with more than two states, deriving and computing exact expressions for the probability of decoding failure rapidly becomes intractable.
In such situations, the use of upper bounds for performance evaluation is inevitable.

\begin{comment}
We begin our survey with a familiar case, the binary symmetric channel.
In this case, the probability of error for maximum-likelihood decoding of a length-$N$ uniform random code with $M$ codewords is given by
\begin{equation*} %\label{equation:ProbErrorBSC}
%P_{\mathrm{e}} =
\sum_{\ell=0}^N \binom{N}{\ell} \varepsilon^{\ell} (1-\varepsilon)^{N-\ell}
\left( 1 - \left( 1 - \frac{1}{2^N} \sum_{j=0}^{\ell} \binom{N}{j} \right)^{M-1}
\right) ,
%\begin{split}
%P_{\mathrm{e}}
%= \sum_{\ell=0}^N & \binom{N}{\ell} \varepsilon^{\ell} (1-\varepsilon)^{N-\ell} \\
%&\times
%\left(1 - \left( 1 - 2^{-N} \sum_{j=0}^{\ell} \binom{N}{j} \right)^{M-1}
%\right) ,
%\end{split}
\end{equation*}
where $\varepsilon$ is the crossover probability of the channel.
The maximum-likelihood decision rule in this case is a minimum-distance decoder, and this error probability can be established through a counting argument.
We note that the derivation of this bound assumes that the decoder treats ties as decoding failures~\cite{Fano-1961}.
It is possible to get slightly better performance by having the receiver select one of the codewords when several candidates are equidistant from the received signal~\cite{Polyanskiy-it10}.
\end{comment}

In~\cite{HamidiSepehr-arxiv13}, the authors study data transmission over a Gilbert-Elliott channel using random coding when the state is known at the receiver.
Two different decoding schemes are considered: a minimum-distance decoder and a maximum-likelihood decision rule.
For the sake of completeness, we briefly review these results.
When channel state information is available at the destination, the empirical distribution of the channel sequence provides enough information to determine the probability of decoding failure.
Using the measure on $N_1$ and the corresponding conditional error probabilities, one can average over all possible types to get the probability of decoding failure,
\begin{align} \label{eq:1}
% SINGLE
P_{\mathrm{e},S_N|S_0}(s_N|s_0) &= \sum_{T \in \mathcal{T}} P_{\mathrm{e} | \type (\mathbf{S})} (T)
\Pr(\type (\mathbf{S}) = T,S_N=s_N|S_0=s_0) .
\end{align}
The conditional probability of decoding failure, given type $T=(n_1,N-n_1)$, is examined further below.
The probability distributions governing different channel types can be found in Lemma~\ref{thm:1}.

Consider the channel realization over the span of a codeword.
Suppose $\mathbf{X}_i$ and $\mathbf{Y}_i$ represent the subvectors of $\mathbf{X}$ and $\mathbf{Y}$ corresponding to time instants when the channel is in state $i$.
We can denote the number of errors that occur in each state using random variables $E_1$ and $E_2$, where $E_i = d_{\mathrm{H}} (\mathbf{X}_i, \mathbf{Y}_i)$ and $d_{\mathrm{H}} (\cdot,\cdot)$ is the Hamming distance.
The conditional error probabilities can then be written as
\begin{equation} \label{eq:2}
\begin{split}
% SINGLE
P_{\mathrm{e}|\type(\mathbf{S})} (T)
= \sum_{e_1 = 0}^{n_1} \sum_{e_2 = 0}^{n_2}
&P_{\mathrm{e} | \type (\mathbf{S}), E_1, E_2} (T, e_1, e_2)
P_{E_1, E_2 | \type(\mathbf{S})} (e_1, e_2 | T) ,
\end{split}
\end{equation}
where $n_2=N-n_1$.
Given the channel type, the numbers of errors in the good and bad states are independent,
\begin{equation}  \label{eq:3-1}
P_{E_1, E_2 | \type(\mathbf{S})} (e_1, e_2 | T)
= P_{E_1 | \type(\mathbf{S})} (e_1 | T)
P_{E_2 | \type(\mathbf{S})} (e_2|T) .
\end{equation}
Furthermore, $E_1$ and $E_2$ have binomial distributions
\begin{equation}\label{eq:3-2}
P_{E_i | \type(\mathbf{S})} (e_i | T)
= P_{E_i | N_i}(e_i | n_i)
= \binom{n_i}{e_i} \varepsilon_i^{e_i} (1 - \varepsilon_i)^{n_i - e_i} .
\end{equation}
This mathematical structure leads to the following theorem.

\begin{theorem} \label{GEC_ML_BER}
When ties are treated as errors, the probability of decoding failure for a length-$N$ uniform random code with $M$ codewords, conditioned on the number of symbol errors in each state and the channel state type, is given by
\begin{equation} \label{eq:7}
\begin{split}
% SINGLE
&P_{\mathrm{e}|\type(\mathbf{S}), E_1,E_2} (T,e_1,e_2)
= 1 - \left(1 - 2^{-N} \sum_{(\tilde{e}_1,\tilde{e}_2) \in \mathcal{M}(\gamma e_{1}+e_{2})} \binom{n_1}{\tilde{e}_1} \binom{n_2}{\tilde{e}_2} \right)^{M - 1}
\end{split}
\end{equation}
where $\mathcal{M}(d)$ is the set of pairs $(\tilde{e}_1, \tilde{e}_2) \in \{0, \ldots, N\}^2$ that satisfy $\gamma \tilde{e}_1 + \tilde{e}_2 \leq d$.
This expression holds with
\begin{equation*}
\gamma = \frac{\ln \varepsilon_1 - \ln(1-\varepsilon_1)}
{\ln \varepsilon_2-\ln(1-\varepsilon_2)}
\end{equation*}
for the maximum-likelihood decision rule, and with $\gamma=1$ for minimum-distance decoding.
\end{theorem}
\begin{IEEEproof}
See \cite{HamidiSepehr-arxiv13}.
\end{IEEEproof}

If the channel state is  causally known at the receiver, then random codes paired with a maximum-likelihood decoding rule form a permutation invariant scheme.
The decoding performance is then determined by the number of symbol errors in each state within a codeword, and not by their order or locations. 
We point out that the methodology introduced herein can potentially be extended to other permutation invariant encoding/decoding schemes to analyze probability of decoding failure.

\section{Undetected Errors}

\begin{comment}
Undetected errors can cause serious problems in real systems and the probability of undetected error is an important performance metric for this reason.
In delay-sensitive applications, one worries about undetected errors even more because detection and recovery times can lead to very large delays.
Therefore, we apply the standard modification of the maximum-likelihood decoding rule to reduce the probability of undetected error~\cite{Forney-it68}.
During the design phase, the probability of undetected error is required to be small enough so that system constraints are satisfied.
\end{comment}

A serious matter with pragmatic communication systems is the presence of undetected decoding failures.
In the current setting, this occurs when the receiver uniquely decodes to the wrong codeword.
For delay-sensitive applications, this problem is especially important because recovery procedures can lead to undue delay.
To address this issue, we apply techniques that help control the probability of admitting erroneous codewords \cite{Forney-it68,Hof-it10}.
This safeguard, in turn, leads to slight modifications to the performance analysis presented above.
In applying these techniques, the probability of undetected failure is a system parameter that must be set during the design phase of the system.

\subsection{The Exact Approach}

In \cite{HamidiSepehr-arxiv13}, the authors show that the probability of decoding failure (including detected errors, undetected errors, and ties) is given by the equations \eqref{eq:1}--\eqref{eq:3-2} and substituting 
\begin{equation} \label{eq:7-nu}
\begin{split}
&{P}_{\mathrm{e}|\type(\mathbf{S}), E_1,E_2} (T,e_1,e_2)
= 1 - \left(1 - 2^{-N} \sum_{(\tilde{e}_1,\tilde{e}_2) \in \mathcal{M}(\gamma e_{1}+e_{2}+\nu)} \binom{n_1}{\tilde{e}_1} \binom{n_2}{\tilde{e}_2} \right)^{M - 1},
\end{split}
\end{equation}
\begin{comment}
by % Why is this not exact?
\begin{align}
& \bar{P}_{\mathrm{e},S_N|S_0}(d|c) = \nonumber \\
&\sum_{n_1=0}^{N} \sum_{e_1=0}^{n_1} \sum_{e_2=0}^{n_2}
\binom{n_1}{e_1} \binom{n_2}{e_2}
\varepsilon_1^{e_1}(1\!-\!\varepsilon_1)^{n_1-e_1}
\varepsilon_2^{e_2}(1\!-\!\varepsilon_2)^{n_2-e_2} \nonumber\\
&\times \left[ 1 - \left(1 - 2^{-N}\!\!\!\!\!\!
\sum_{(\tilde{e}_{1},\tilde{e}_{2}) \in \mathcal{M} (\gamma e_{1}+e_{2}+\nu)}
\binom{n_{1}}{\tilde{e}_{1}} \binom{n_{2}}{\tilde{e}_{2}} \right)^{M-1} \right]\nonumber\\
&\times P_{N_1,S_N|S_0}(n_1,d|c) \label{eq:err}
\end{align}
\end{comment}
where $\nu$ is a non-negative parameter that specifies the size of the safety margin for undetected errors.
%
\begin{comment}
A \emph{detected error} occurs when the decoder chooses not to output a codeword.
The problem is setup so that the correct codeword is always on the list.
Therefore, an undetected error may occur only if there is another codeword inside the ball of radius $C-\nu$.
If there are only two codewords in the ball of radius $C$ and one of them is inside the ball of radius $C-\nu$, then the decoder will return the incorrect codeword.
On the other hand, it is possible that two incorrect codewords may fall inside the ball of radius $C-\nu$ and still be close enough together to cause a detected error.
Though, the probability of such events is very small.
\end{comment}
%
The joint probability of undetected error with ending state $S_N$, conditioned on starting in state $S_0$, is $P_{\mathrm{ue},S_N|S_0}(s_N|s_0)$ and can be upper bounded by
\begin{equation} \label{eq:1_ue}
\begin{split}
% SINGLE
&\bar{P}_{\mathrm{ue},S_N|S_0}(s_N|s_0)
= \sum_{T \in \mathcal{T}} \bar{P}_{\mathrm{ue} | \type (\mathbf{S})} (T)
\Pr(\type (\mathbf{S}) = T,S_N=s_N|S_0=s_0)
\end{split}
\end{equation}
where 
\begin{equation} \label{eq:2_ue}
\begin{split}
% SINGLE
\bar{P}_{\mathrm{ue}|\type(\mathbf{S})} (T)
= \sum_{e_1 = 0}^{n_1} \sum_{e_2 = 0}^{n_2}
&\bar{P}_{\mathrm{ue} | \type (\mathbf{S}), E_1, E_2} (T, e_1, e_2)
P_{E_1, E_2 | \type(\mathbf{S})} (e_1, e_2 | T) ,
\end{split}
\end{equation}
and
\begin{equation} \label{eq:7-nu_ue}
\begin{split}
% SINGLE
&\bar{P}_{\mathrm{ue}|\type(\mathbf{S}), E_1,E_2} (T,e_1,e_2)
= 1 - \left(1 - 2^{-N} \sum_{(\tilde{e}_1,\tilde{e}_2) \in \mathcal{M}(\gamma e_{1}+e_{2}-\nu)} \binom{n_1}{\tilde{e}_1} \binom{n_2}{\tilde{e}_2} \right)^{M - 1}.
\end{split}
\end{equation}
%\vspace{-2mm}
\begin{comment}
\begin{align}
&\bar{P}_{\mathrm{ue},S_N|S_0}(d|c) =\nonumber\\
&\sum_{n_1=0}^{N} \sum_{e_1=0}^{n_1} \sum_{e_2=0}^{n_2}
\binom{n_1}{e_1} \binom{n_2}{e_2}
\varepsilon_1^{e_1}(1\!-\!\varepsilon_1)^{n_1-e_1}
\varepsilon_2^{e_2}(1\!-\!\varepsilon_2)^{n_2-e_2} \nonumber\\
&\times \left[ 1 - \left(1-2^{-N}\!\!\!\!\!\!\sum_{(\tilde{e}_{1},\tilde{e}_{2})\in\mathcal{M}(\gamma e_{1}+e_{2}-\nu)}\binom{n_{1}}{\tilde{e}_{1}}\binom{n_{2}}{\tilde{e}_{2}}\right)^{M-1}\right]\nonumber\\
&\times P_{N_1,S_N|S_0}(n_1,d|c).\label{eq:undet}
\end{align}
\end{comment}
%% COMMENT
\begin{comment}
Notice that ties are treated as detected errors. This follows naturally from choosing $\nu=0$.
In Theorem~\ref{GEC_ML_BER}, there is a single ball around the received vector and a tie is a special case that is treated as a failure.
For $\nu=0$, we now have only one term in the upper bound on $P_{\mathrm{e},S_N|S_0}(d|c)$ and ties becomes detected errors.

It is clear that the probability of undetected error is decreasing in $\nu$ and it can be shown to decay
exponentially in $\nu$. Therefore, by choosing an appropriate value for $\nu$, we can make the the decoder more
robust and force the undetected error to be as small
as the application requires.
\end{comment}
%% COMMENT
Since the probability of an undetected error is typically much smaller than that of a detected error, one can upper bound the probability of detected error by $P_{\mathrm{e},S_N|S_0}(s_N|s_0)$ with a negligible penalty.

\subsection{Exponential Bound}

With slight modifications to the derived exponential upper bound, one can get a similar bound on the probability of undetected error.
\begin{lemma}
The exponential upper bounds on ${P}_{\mathrm{e},S_N|S_0}$, and $\bar{P}_{\mathrm{ue},S_N|S_0}$ can be written as\vspace{-2mm}
\begin{align}
% SINGLE
&\tilde{P}_{\mathrm{ue}, S_N | S_0} (j  | i)
= \min_{0\leq v\leq\rho\leq1}
\sum_{n_1=0}^N\!\min \Big\{\! 1,e^{- N ( E_{0,N} (\rho, \mathbf{Q}_N, n_1) - \rho R -v\tau) } \!\Big\} P_{N_1,S_N|S_0}(\!n_1,j  | i) \nonumber\\
& \approx\!\min_{0\leq v\leq\rho\leq1} \int_0^1 \!\! \min \Big\{ 1, e^{ - N ( (b_1(\rho)-b_2(\rho)) r + b_2(\rho) - \rho R -v\tau) } \Big\}
f_{\eta, S_{\mathrm{f}} | S_{\mathrm{i}}} (r, j  | i) dr \label{eq:undet_bound}
\end{align}
and
\begin{align}
% SINGLE
&\tilde{P}_{\mathrm{e}, S_N | S_0} (j  | i)
= \min_{0\leq v\leq\rho\leq1}
\sum_{n_1=0}^N\!\min \left\{\! 1,e^{- N ( E_{0,N} (\rho, \mathbf{Q}_N, n_1) - \rho R -v\tau+\tau) } \!\right\}\! P_{N_1,S_N|S_0}(\!n_1,j  | i) \nonumber\\
& \approx\!\min_{0\leq v\leq\rho\leq1} \int_0^1 \!\! \min \left\{\! 1, e^{ - N ( (b_1(\rho)-b_2(\rho)) r + b_2(\rho) - \rho R -v\tau+\tau) }\! \right\}
f_{\eta, S_{\mathrm{f}} | S_{\mathrm{i}}} (r, j  | i) dr \label{eq:err_bound}
\end{align}
where $\tau$ controls the tradeoff between detected and undetected errors and is used to decrease the incidence of undetected errors, in a manner similar to $\nu$ for the exact case.
\end{lemma}
\begin{IEEEproof}
Following the same approach as in \cite{Forney-it68,Hof-it10}, the results are achieved.
\end{IEEEproof}

We emphasize that the rare-transition regime, the bounds and the approximation methodologies proposed in this article have a wide range of applications. 
In fact, the introduced upper bounds can be adopted in various analysis frameworks.
The following two sections are dedicated to the potential implications of the proposed bounding techniques in terms of queueing theory. 
We exploit these results to evaluate the queueing performance of systems built around correlated channels. 
In particular, we show how stochastic dominance can be combined with these tools to make performance analysis tractable.

\section{Queueing Model}
\label{sec:queue}

Consider a queueing system in which packets are generated at the source according to a Poisson process with arrival rate $\lambda$, measured in packets per channel use.
The number of bits per packet forms a sequence of independent geometric random variables, each with parameter $\rho \in (0,1)$.
On arrival, a packet is divided into segments of size $RN / \ln 2$~bits, where $N$ denotes the block length and $R$ is the code rate in nats per code bit, as defined in Section~\ref{section:Modeling}.
%In other words, $N$ represents the number of code bits per segment.
The number of information bits per segment, $RN / \ln 2$, is assumed to be an integer.
Since the convention for rate differs between the exponential upper bounds and the standard coding theory, we let $\tilde{R}={R}/{\ln 2}$ be the code rate in information bits per code bit associated with the code rate $R$ in information nats per code bit.
The total number of segments associated with a packet of length $L$ is given by $J = \left\lceil \frac{L }{\tilde{R}N} \right \rceil$.
As before, a random coding scheme is used to protect the transmitted data while it is transmitted over the correlated channel.

On the receiving end, one must successfully decode all $J$ codewords to recover the corresponding packet.
Once this is achieved, this packet is discarded from the queue.
We note that random variable $J$ has a geometric distribution with
$\Pr(J=j) = (1 - \rho_r)^{j-1} \rho_r$, where $j \geq 1$ and $\rho_r = 1 - (1 - \rho)^{\tilde{R}N }$.
Consequently, the number of coded blocks per data packet possesses the memoryless property, a highly desirable attribute for the purpose of analysis.

This queueing system operates on top of the finite-state channel discussed earlier.
In comparison to many previous studies, the resulting framework allows us to rigorously characterize the queueing performance while varying the block length and the code rate.
%The assumption of Poisson packet arrivals with geometric packet length, allows us to scale the arrival process
The scaling property of the Poisson packet arrivals with geometric packet lengths is crucial in enabling the fair comparison of systems with different code parameters.
In particular, the arrival process is defined at the bit level and we account for channel dependence within and across codewords.
This last observation is especially pertinent for queueing systems, as correlation in service is known to exacerbate the distribution of a queue.

We emphasize that, in the current framework, a data packet is discarded from the transmit buffer if and only if the destination acknowledges reception of the latest codeword and this codeword contains the last parcel of information corresponding to the head packet.
Packet departures are then determined by the channel realizations and the coding scheme.
In particular, the code rate $\tilde{R}$ has a major impact on performance.
Generally, a lower code rate will have a small probability of decoding failure.
However, it also needs more channel uses to complete the transmission of one data packet.
Thus, for a fixed channel profile, we can vary the block length and code rate to find optimal system parameters.
This natural tradeoff reflects the tension between the probability of a successful transmission and the size of its payload.

Let $Q_{s}$ denote the number of data packets waiting in the transmitter queue after $s$ codeword transmission intervals.
The channel state at the same time instant is represented by $C_{sN+1}$.
Notice that the channel state evolves more rapidly than events taking place in the queue.
This explains the discrepancy between the indices.
Based on these quantities, it is possible to define a Markov chain $U_s = (C_{sN+1},{Q}_{s})$ that captures the joint evolution of the queue and the channel over time.
The ensuing transition probabilities from $U_s$ to $U_{s+1}$ are equal to
\begin{equation*}
\begin{split}
% SINGLE
\Pr(U_{s+1} = (d,q_{s+1}) | U_s = (c,q_s)) =
&\sum_{n_1=0}^{N} P_{{Q}_{s+1}|N_1,Q_s} (q_{s+1}|n_1,q_s)
P_{N_1,C_{(s+1)N+1}|C_{sN+1}}(n_1,d|c),
\end{split}
%\label{equation:StateTranstionProbabilities}
\end{equation*}
where the second term in the summand is given in Lemma~\ref{thm:1}.
We can rewrite $P_{Q_{s+1}|N_1,Q_s} \left( q_{s+1}|n_1,q_s \right)$ as
\begin{equation*}
\begin{split}
% SINGLE
&P_{Q_{s+1}|N_1,Q_s} \left( q_{s+1}|n_1,q_s \right)
= \sum_{e_1=0}^{n_1} \sum_{e_2=0}^{n_2}
P_{Q_{s+1},E_1,E_2|N_1,Q_s} (q_{s+1},e_1,e_2|n_1,q_s) \\
&= \sum_{e_1=0}^{n_1} \sum_{e_2=0}^{n_2}
P_{Q_{s+1}|E_1,E_2,N_1,Q_s}(q_{s+1}|e_1,e_2,n_1,q_s)
P_{E_1,E_2|N_1,{Q}_s}(e_1,e_2|n_1,q_s) \\
&= \sum_{e_1=0}^{n_1} \sum_{e_2=0}^{n_2} \binom{n_1}{e_1} \binom{n_2}{e_2}
\varepsilon_1^{e_1}(1-\varepsilon_1)^{n_1-e_1}
\varepsilon_2^{e_2}(1-\varepsilon_2)^{n_2-e_2}
P_{Q_{s+1}|E_1,E_2,N_1,Q_s}(q_{s+1}|e_1,e_2,n_1,q_s) .
\end{split}
%\label{equation:StateTranstionProbabilities}
\end{equation*}
Suppose that the number of packets in the queue is $Q_s = q_s$, where $q_s > 0$.
Then, admissible values for $Q_{s+1}$ are restricted to the set $\{ q_s-1, q_s, q_s+1, \ldots\}$.
The transition probabilities for $q_s > 0$ and $i \geq 0$ are given by
\begin{equation} \label{equation:QueueIncrease}
\begin{split}
% SINGLE
&P_{{Q}_{s+1}|E_1,E_2,N_1,Q_s} (q_s+i|e_1,e_2,n_1,q_s) \\
&\quad = a_i {P}_{\mathrm{e}|E_1,E_2,N_1}(e_1,e_2,n_1)
+ a_i (1-{P}_{\mathrm{e}|E_1,E_2,N_1}(e_1,e_2,n_1)) (1 - \rho_r) \\
&\quad + a_{i+1} (1 - {P}_{\mathrm{e}|E_1,E_2,N_1}(e_1,e_2,n_1)) \rho_r
\end{split}
\end{equation}
and the probability of the queue decreasing is
\begin{equation} \label{equation:QueueDecrease}
\begin{split}
% SINGLE
&P_{Q_{s+1}|E_1,E_2,N_1,Q_s}(q_s - 1|e_1,e_2,n_1,q_s)
= a_0 \left( 1 - {P}_{\mathrm{e}|E_1,E_2,N_1} (e_1,e_2,n_1) \right) \rho_r .
\end{split}
\end{equation}
The queue can only become smaller when there are no arrivals, a codeword is successfully received at the destination, and the decoded codeword contains the last piece of data associated with a packet.
Above, ${P}_{\mathrm{e}|E_1,E_2,N_1}(e_1,e_2,n_1)$ is the conditional probability of decoding failure which appears in \eqref{eq:7-nu}. %; it appears as part of the summand in \eqref{eq:err}.
The terms $a_i$ denotes the probability that $i$ packets arrive within the span of a codeword transmission.
Since arrivals form a Poisson process, we have $a_i = \frac{(\lambda N)^i}{i!} e^{-\lambda N}$ for $i\geq0$.
When the queue is empty, $q_s = 0$, \eqref{equation:QueueIncrease} applies for cases where $i \geq 1$.
However, for $i = 0$, the conditional transition probability reduces to
\begin{equation*}
\begin{split}
% SINGLE
&P_{Q_{s+1}|E_1,E_2,N_1,Q_s}(0|e_1,e_2,n_1,0)
= a_0 + a_1
\left( 1 - {P}_{\mathrm{e}|E_1,E_2,N_1} (e_1,e_2,n_1) \right) \rho_r .
\end{split}
\end{equation*}

\begin{comment}
Next, we can get the probability transition matrix of the Markov process $\{\tilde{U}_{s}\}$.
For convenience, first we introduce the following mathematical notation,
\begin{align}
\mu_{cd}^{i} & =\Pr(U_{s+1}=(d,q+i)|U_{s}=(c,q))\,,\qquad i\ge1\label{eq:mu2}\\
\kappa_{cd} & =\Pr(U_{s+1}=(d,q)|U_{s}=(c,q))\label{eq:kappa}\\
\xi_{cd} & =\Pr(U_{s+1}=(d,q-1)|U_{s}=(c,q)).\label{eq:zeta}
\end{align}
Similarly, when the queue is empty, we have
\begin{align}
\mu_{cd}^{i0} & =\Pr(U_{s+1}=(d,i)|U_{s}=(c,0))\label{eq:mu0}\\
\kappa_{cd}^{0} & =\Pr(U_{s+1}=(d,0)|U_{s}=(c,0)),\label{eq:kappa0}
\end{align}
where $q\in\mathbb{N}_{0}$ and $c,d\in\{1,2\}$.
\end{comment}

The overall profile of this system can be categorized as an M/G/1-type queue.
The repetitive structure enables us to employ the matrix geometric method to compute the characteristics of this system and subsequently obtain its stationary distribution \cite{Riska-sigmet02,HamidiSepehr-arxiv13}.

\section{Stochastic Dominance}
\label{sec:Stochastic-Dominance}

When the number of channel states is large, it may be impractical to employ exact probabilities of decoding failure.
Even for memoryless channels, finding explicit expressions for different encoding/decoding schemes can be difficult.
In the face of such a challenge, it is customary to turn to upper bounds on the probabilities of decoding failure to provide performance guarantees.
Furthermore, one can employ such upper bounds to assess the queueing performance of the system through stochastic dominance.
We emphasize that this type of argument is independent of our proposed bounds.
Rather, it applies to any upper bound on the probability of decoding failure.

The evolution of the queue length is governed by the Lindley equation,
\begin{equation} \label{Q_s}
\begin{split}
% SINGLE
Q_{s+1} &= \left( Q_s + A_s - D_s \right)^+
\triangleq \max \{ 0, Q_s + A_s - D_s \}
\end{split}
\end{equation}
where $A_s$ is the number of arrivals that occurred during time interval~$s$, and $D_s$ is an indicator function for the potential completion of a packet transmission within the same time period.
In this queueing model, the only inherent effect of replacing the probability of decoding failure by an upper bound is a potential reduction in the value of $D_s$.
Using an upper bound on the failure probability naturally gives rise to a new random process $\tilde{Q}_s$ defined by
\begin{equation}
\label{Qtilde_s}
\tilde{Q}_{s+1} = \left( \tilde{Q}_s + A_s - \tilde{D}_s \right)^+
\end{equation}
where $\tilde{D}_s$ is drawn according to the distribution implied by the upper bound.
We wish to show that, conditioned on starting in the same state and given a shared channel trace, the distribution of $\tilde{Q}_s$ offers a conservative estimate of $Q_s$.
To make this statement precise, we turn to an establish concept in probability.

\begin{definition}
A random variable $Z$ is stochastically dominated by another random variable $\tilde{Z}$, a relation which we denote by $Z \preceq \tilde{Z}$, provided that
\begin{equation} \label{stochasticOrder}
\Pr (Z > z) \leq \Pr (\tilde{Z} > z)
\end{equation}
for all $z \in \mathbb{R}$.
This relation extends to conditional probability laws.
Suppose that
\begin{equation} \label{equation:ConditionalStochasticOrder}
\Pr (Z > z | A) \leq \Pr (\tilde{Z} > z | A)
\end{equation}
for all $z \in \mathbb{R}$.
Then, we say that $Z$ is stochastically dominated by $\tilde{Z}$, given $A$.
We write this relation as $Z\preceq_A \tilde{Z}$.
\end{definition}

A comprehensive discussion of stochastic dominance can be found in \cite{Shaked-06,Stoyan-83}.
In order to formulate the results we are interested in, we need to start by introducing two lemmas.

\begin{lemma} \label{prop:positivePart}
The stochastic order defined in \eqref{stochasticOrder} is preserved under the positive part operation, $(\cdot)^+ = \max \{0, \cdot\}$.
\end{lemma}
\begin{IEEEproof}
From \cite[Sec.~1.A.1]{Shaked-06}, we know that given random variables $X$ and $Y$, $X\preceq Y$ if and only if 
\begin{equation*}
\mathbb{E}[\varphi(X)]\leq \mathbb{E}[\varphi(Y)]
\end{equation*} 
for all increasing functions $\varphi(\cdot)$ for which the expectations exists.
In particular, $\varphi(\cdot) = \max\{0, \cdot\}$ is an increasing function.
Thus, if $X \preceq Y$ then one can conclude that $X^+\preceq Y^+$, as desired.
\end{IEEEproof}

We explore the structure of potential packet completion events below.
In establishing stochastic dominance, we will condition on a specific channel trace,
\begin{equation*}
\vec{C} = \{ C_1, C_{N + 1}, C_{2N + 1}, \ldots \}
\end{equation*}
This ensures that the two stochastic processes experience a same level of difficulty at every step while trying to decode codewords.
Note that, for our purposes, the dependence of $Q_s$ and $\tilde{Q}_s$ on $\vec{C}$ is only through decoding attempts and, as such, this dependence is localized in time.

\begin{lemma} \label{lemma:DominanceD}
Using upper bounds on the probabilities of decoding failure leads to stochastic dominance in potential packet completions, $\tilde{D}_s \preceq_{\vec{C}} D_s$.
\end{lemma}
\begin{IEEEproof}
A potential completion occurs when a codeword is decoded successfully at the destination, and the data it contains is the last segment of a packet.
The probability of the latter event is $\rho_r$, and it is common to both $D_s$ and $\tilde{D}_s$.
However, the conditional probabilities of decoding failure differ, with $P_{\mathrm{e}|\vec{C} = \vec{c}} \leq \tilde{P}_{\mathrm{e}|\vec{C} = \vec{c}}$.
This, in turn, gives
\begin{equation*}
\begin{split}
% SINGLE
&\Pr \left( \tilde{D}_s = 1 | \vec{C} = \vec{c} \right)
= \left( 1 - \tilde{P}_{\mathrm{e} | \vec{C} = \vec{c}} \right) \rho_r
\leq \left( 1 - P_{\mathrm{e}| \vec{C} = \vec{c}} \right) \rho_r
= \Pr \left( D_s = 1 | \vec{C} = \vec{c} \right) .
\end{split}
\end{equation*}
Since $D_s, \tilde{D}_s \in \{0, 1\}$, this equation is enough to establish stochastic ordering.
\end{IEEEproof}

Collecting these results, we can turn to the behavior of the system.
For a fair comparison, we assume that the two queues have the same number of packets at the onset of the communication process.

\begin{proposition} \label{lemm:stochDom}
Suppose that $Q_0 = \tilde{Q}_0$.
Then, the process $Q_s$ is stochastically dominated by $\tilde{Q}_s$.
\end{proposition}
\begin{IEEEproof}
As a first step, we assume that channel trace $\vec{C} = \vec{c}$ is fixed.
We prove conditional dominance through mathematical induction.
The base case follows from the condition of the theorem.
As an inductive hypothesis, suppose that $Q_s \preceq_{\vec{C}} \tilde{Q}_s$.
By Lemma~\ref{lemma:DominanceD}, we have $\tilde{D}_s \preceq_{\vec{C}} D_s$.
Since negation reverses stochastic dominance~\cite[pp.~9]{Shaked-06}, we deduce that
\begin{equation*}
- D_s \preceq_{\vec{C}} -\tilde{D}_s .
\end{equation*}
This, in turn, yields the relation
\begin{equation*}
Q_s + A_s - D_s  \preceq_{\vec{C}} \tilde{Q}_s + A_s - \tilde{D}_s .
\end{equation*}
The last step leverages the closure property of stochastic orders under convolutions \cite[Thm.~1.A.3(b)]{Shaked-06}.
Applying the increasing function $\varphi(\cdot) = \max\{0, \cdot\}$ to both sides, we immediately get
\begin{equation*}
\left( Q_s + A_s - D_s \right)^+ \preceq_{\vec{C}} \left( \tilde{Q}_s + A_s - \tilde{D}_s \right)^+
\end{equation*}
from Lemma~\ref{prop:positivePart}.
That is, $Q_{s+1} \preceq_{\vec{C}} \tilde{Q}_{s+1}$, thereby establishing our inductive step.

At this point, the statement of the proposition can be obtained by taking expectations over channel traces.
Since the channel states are independent of queue sizes and code generation, the probabilistic weighing is the same for both $Q_s$ and $\tilde{Q}_s$.
This guarantees that the stochastic ordering is preserved.
In other words, the relation $Q_s \preceq \tilde{Q}_s$ holds at all times.
\end{IEEEproof}

There is a subtle distinction in the argument presented above.
The random variable $D_s$ indicates potential completion.
Actual departures from the queue only take place when there are packets awaiting transmission.
Mathematically, the distinction is resolved through the positive part operation.
Conceptually, when the queue is empty, the source attempts to send a virtual packet with no physical meaning.
This object is created for mathematical convenience.

In view of the aforementioned results, we can also consider the interplay between stochastic dominance and the stationary distributions of the queues.
When the Markov chains $U_s$ and $\tilde{U}_s$ are positive recurrent, the corresponding queueing processes $Q_s$ and $\tilde{Q}_s$ are stable.
In such cases, the stationary distribution associated with $\tilde{Q}_s$ dominates the stationary distribution of $Q_s$.
In particular, for any integer $q$, we have $\Pr (Q_s > q) \leq \Pr \left( \tilde{Q}_s > q \right)$ and, hence, in the limit we obtain
\begin{equation*}
\begin{split}
\Pr (Q > q) &= \lim_{s \rightarrow \infty} \Pr (Q_s > q) \\
&\leq \lim_{s \rightarrow \infty} \Pr \left( \tilde{Q}_s > q \right)
= \Pr \left( \tilde{Q} > q \right)
\end{split}
\end{equation*}
That is, $Q \preceq \tilde{Q}$ where $Q$ and $\tilde{Q}$ denote the stationary distributions of the two queueing processes listed above.

From a more intuitive point of view, one can argue that, pathwise, increasing the probability of failure can only result in fewer departures and, as such, there will remain at least as many packets waiting in the queue.
In other words, when comparing two queueing systems with a same arrival process, a same underlying channel, and a same code generator, more decoding failures can only exacerbate the size of the queue.
%So when analyzing the system based on the upper bound on error probability, one can be sure that the queueing performance of the original system will be no worse. The actual average queue length and consequently, the actual average delay will be lower than those of the estimated system. This means that if the estimated system satisfies certain criteria, the original system also satisfies them.
This observation holds in some generality and can be employed when the exact decoding error probability are not known or difficult to compute.
%Even for the two-state Gilbert-Elliott channel, the computation of the exact error probability is computationally challenging for block lengths greater than 150.
This approach allows one to provide performance guarantees for a queueing system using bounds on the probabilities of decoding failure.

\section{Numerical Results}
\label{sec:numerical}

In this section, we present numerical results for probabilities of decoding error and we compare them to the derived upper bounds.
We also evaluate queueing performance using the exact error probabilities and their upper bounds derived in the rare-transition regime.

\subsection{Comparison of Exponential Upper Bounds}

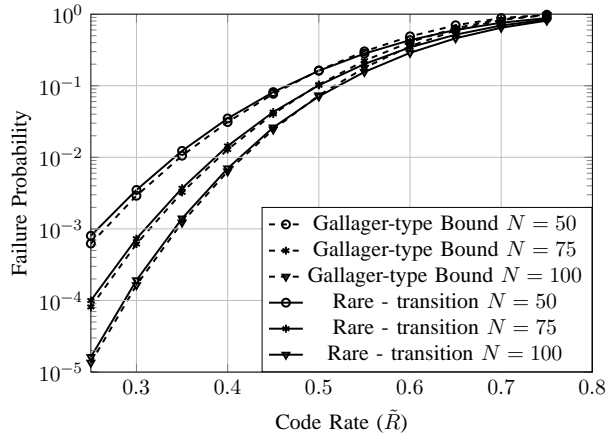
\begin{figure}
\begin{center}
\input{Figures/bound_plots_avgres}
\end{center}
\caption{Comparison of the approximate upper bound \eqref{equation:ApproximateUpperBound} with the exact bound \eqref{equation:ErrorBound} in the rare-transition regime with $N[\mathbf{P}_N]_{12}\approx4$ and $N[\mathbf{P}_N]_{21}\approx6$.}
\label{figure:bounds}
\end{figure}

We consider a communication system which transmits data over a Gilbert-Elliott channel with crossover probabilities $\varepsilon_1=0.01$ and $\varepsilon_2=0.1$.
Figure~\ref{figure:bounds} shows the approximate upper bounds of \eqref{equation:ApproximateUpperBound} as functions of block length and code rate, and compares them to the standard Gallager-type bounds of \eqref{equation:ErrorBound}.
%The rare-transition regime is characterized by
%\begin{equation*}
%\mathbf{P}_N = \begin{bmatrix}
%1 - \frac{\alpha}{N} & \frac{\alpha}{N} \\
%\frac{\beta}{N} & 1 - \frac{\beta}{N}
%\end{bmatrix}
%\end{equation*}
%where $\alpha=0.04$ and $\beta=0.12$.
Each curve shows the value of the bound averaged over all possible state transitions.
Although the block lengths are relatively short, the approximate bounds are very close to the standard Gallager-type bounds.
Furthermore, the difference becomes more negligible as $N$ grows larger.

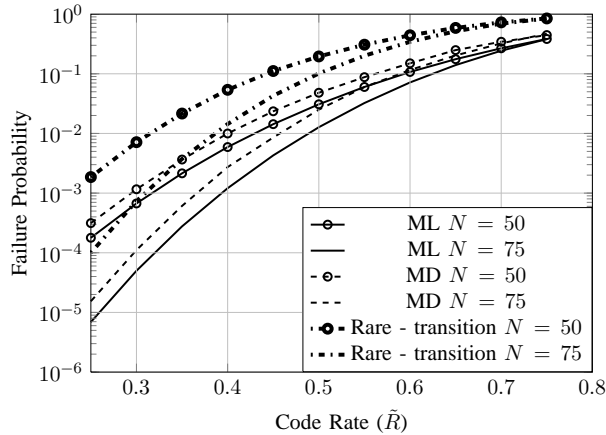
\begin{figure}
\begin{center}
{\input{Figures/boundvsexacts_plots_avgres}}
\end{center}
\caption{Comparison of the approximate upper bound \eqref{equation:ApproximateUpperBound} with the exact probabilities of decoding failure under maximum-likelihood (ML) and minimum-distance (MD) decoding for {$[\mathbf{P}_N]_{12}= 0.0533$} and {$[\mathbf{P}_N]_{21} = 0.08$}.}
\label{figure:exactVSbounds}
\end{figure}

In Fig.~\ref{figure:exactVSbounds}, we plot the probabilities of decoding failure for the maximum-likelihood and minimum-distance decoders given by \eqref{eq:1}-\eqref{eq:7}, against the bounds provided in \eqref{equation:ApproximateUpperBound}.
As anticipated, the maximum-likelihood decision rule outperforms the minimum distance decoder.
For fixed $N$, there is a roughly constant ratio between the approximate upper bounds and the exact probabilities of error under maximum-likelihood decoding.
This is not too surprising as similar statements can be made about the accuracy of Gallager-type bounds.
%Still, as block length increases, the approximate bounds get progressively closer to the exact values.
We note that the figure features particularly short block lengths, as it is impractical to compute exact performance for long lengths.

\subsection{Evaluation of Queueing Performance}

We turn to the evaluation of overall performance and we consider a situation where, on average, packets are generated every $20$ msec.
This yields a rate of $\lambda=50$ packets per second for the arrival process.
The symbol rate for our binary channel is set to $28.75$~Kb per second, which leads to an expected ${1}/{575}$ packets per channel use.
The cross-over probabilities of the Gilbert-Elliott channel are set to $\varepsilon_1=0.01$ and $\varepsilon_2=0.1$, and its state transition probabilities are $\alpha=0.0533$ and $\beta=0.08$.
Shannon capacity when the state is known at the receiver is therefore equal to $0.764$ bits per channel use.

Increasing code rate $\tilde{R}$ for a fixed block length decreases redundancy and therefore reduces the error-correcting capability of the code.
Thus, the probability of decoding failure becomes larger.
At the same time, changes in code rate affect $\rho_r$, the probability with which a codeword contains the last parcel of information of a packet.
As code rate varies, these two phenomena alter the transition probabilities and, hence, they influence the stationary distribution of the Markov system in opposite ways.

The choice of a Poisson arrival process allows us to make fair comparisons between codes with different block lengths.
In particular, the rate $\lambda$ in packets per channel use is fixed, and arrivals in the queue correspond to the number of packets produced by the source during the transmission time of one codeword.
The marginal distribution of the sampled process is Poisson with arrival rate $\lambda N$, in packets per codeword.
This formulation is new, and it bridges coding decision to queueing behavior in a rigorous manner.
To examine overall system performance, we assume the existence of a genie which informs the receiver when an undetected decoding error occurs.
Undetected errors are intrinsic to error channels, and this approach is standard when it comes to analysis.
Still, for consistency, we require the system to feature a very low probability of undetected error, e.g., less than $10^{-5}$, by a proper choice of the safety margin.
That is, we only consider $(N,\tilde{R})$ pairs that meet this additional constraint.

Given this framework, a prime goal is to minimize the tail probability of the queue over all admissible values of $N$, $\tilde{R}$, and $\tau$ or $\nu$ that satisfy the constraint on undetected error.
To perform this task using the approximate exponential bound, we first evaluate the bound on undetected error probabilities for different rates and for $\tau=0$ in \eqref{eq:undet_bound}.
Then, for rates with high probability of undetected error, we increase $\tau$ so that the bound on probability of undetected error is decreased.
Recall that this increases the probability of decoding failure. % (see \eqref{eq:undet_bound}--\eqref{eq:err_bound}).
As we are also interested in minimizing the latter probability, we increase $\tau$ until the system meets the error-detecting condition and then stop.
The values of $N$ and $\tilde{R}$ for which this procedure gives poor performance are ignored.
A similar approach is used for system evaluation with exact error probabilities by changing the value of $\nu$ in \eqref{eq:7-nu}, \eqref{eq:7-nu_ue}.

Figure~\ref{fig:tail_bound} shows the approximate probability of the queue exceeding a threshold as a function of system parameters.
The constraint on the number of packets in the queue is set to five, which reflects our emphasis on delay-sensitive communication.
We have chosen $\tau$ in \eqref{eq:undet_bound}--\eqref{eq:err_bound} such that
$\max_{i,j} \tilde{P}_{\mathrm{ue},S_N|S_0}(j|i)$
remains below $10^{-5}$.
The code rate considered vary from $0.25$ to $0.75$, with a step size of $0.05$.
Each curve corresponds to a different block length.
As seen on the graph, there is a natural tradeoff between the probability of decoding failure and the payload per codeword.
For a fixed block length, neither the smallest segment length nor the largest one delivers optimal performance.
Moreover, block length must be selected carefully; longer codewords do not necessarily yield better queueing performance as they may result in large decoding delays.
As such, the tail probability has a minimum over all rates and block lengths.
Therefore, there are interior optimum points for both $N$ and $\tilde{R}$.
We see in Fig.~\ref{fig:tail_bound} that the optimum code parameters are close to $(N,\tilde{R})=(170,0.5)$.
For this particular set of code parameters, we have $\tau=0.048$.

Figure~\ref{fig:tail_exact} offers similar plots for the exact failure probability.
Again, the optimum code parameters are near $(N,\tilde{R})=(170,0.5)$.
In this case, $\nu=8$ is the smallest value of $\nu$ that keeps $\max_{i,j} \bar{P}_{\mathrm{ue},S_N|S_0}(j|i)$ below the $10^{-5}$ threshold.
As we can see by comparing the results, performance evaluation based on the bound gives very good estimates for optimum coding parameters and overall system performance.
Not only does the approximate bound give a good estimate of performance, it accurately predicts ideal system parameters for code block as small as 125.
In addition, since the approximate bounds are slightly pessimistic, they produce conservative estimate of overall performance.   
Empirically, the systems perform better than predicted by the approximate error bounds.

%Although we have an exact expression for the decoding failure probability over the two-state Gilbert-Elliott channel, still the computation time makes it impractical to be used due to the multiple summation and range of $N$, $R$, $\nu$ considered.
%This problem gets worse by increasing the number of channel states.
%In fact, for the channel with more than two states, the exact expressions for the probability of decoding failure are either not available or computationally intractable.
%The proposed approximate bounding technique, provides a convenient framework to design a system with optimum parameters without requiring intractable computations of error probabilities.

\begin{figure}
\begin{center}
 \scalebox{0.65}{\input{Figures/bound2.tex}}
\end{center}
\caption{Approximate bounds on the probability of the queue exceeding a threshold as functions of block length $N$ and code rate $\tilde{R}$.
The system parameters considered above are subject to $\max_{i,j} \tilde{P}_{\mathrm{ue},S_N|S_0}(j|i)\leq10^{-5}$.\label{fig:tail_bound}}

%\end{figure}
%
%\begin{figure}
\vspace{\floatsep}
\begin{center}
 \scalebox{0.65}{\input{Figures/exact.tex}}
\end{center}
\caption{Exact probability of the queue exceeding a threshold as functions of block length $N$ and code rate $\tilde{R}$.
The system parameters presented above meet the constraint $\max_{i,j} \bar{P}_{\mathrm{ue},S_N|S_0}(j|i)\leq10^{-5}$.\label{fig:tail_exact}}
\end{figure}

\section{Conclusion}

The rare-transition regime is a powerful methodology to characterize communication systems where the block length is of the same order or smaller than the coherence time of the channel.
This mode of operation is common in many practical implementations.
This fact serves as a motivation for the proposed framework.
In this article, we derived an approximate upper bound specifically tailored to the rare-transition regime to estimate the probability of decoding error over finite-state channels.
A key property of the proposed methodology is that channel dependence within and across codewords is accounted for in the analysis.
Furthermore, the proposed bound is numerically efficient to compute.
It can be employed for parameter selection and performance analysis in communication links subject to queueing constraints.

We provided supportive evidence for the accuracy of the bounding methodology by deriving exact expressions for the Gilbert-Elliott channel.
Both maximum-likelihood decoding and a minimum-distance decision rule were considered.
A numerical comparison between exact and approximate results validates our approach, showcasing the predictive power of the approximate bounding techniques.
The numerical study focused on a two-state Markov channel formulation where state information is available at the destination.
% Where whould we mention that this results are for the case where the state information is provided and say that in this regime (rare-transition), helps to get similar results for the case where the state inf. is not available at the receiver?
The methodology was subsequently extended to performance criteria based on the queueing behavior of the system.
We provided a practical method to choose the block length and code rate as to minimize the probability that the transmit buffer exceeds a prescribed threshold.
This is especially pertinent for communication links that support delay-sensitive traffic, yet it applies to general data stream as delay is known to negatively affect the performance of flow control and congestion control protocols.
Numerical studies suggest that, for fixed conditions, optimal system parameters are essentially unaffected by small variations in the buffer overflow threshold. 

%In many coded communication systems, the exact decoding failure probability is not known or very complicated to compute.
%Therefore, a stochastic dominance argument is also used to compare the performance of the system computed using an upper bound on decoding failure probability with the exact performance of the system.
%For such systems, our results imply that upper bounds on the decoding failure probability naturally imply upper bounds on the performance of the queueing system.
%Finally, our numerical results imply that, for random coding on the Gilbert-Elliott channel, the performance analysis using the bound on failure probability gives a good estimate of both the system performance and the optimum code parameters.

The methodology and results are developed for finite-state Markov channels, but can be generalized to more intricate channels with memory, with or without symmetry property. 
In addition, the performance characterization of random codes over finite-state channels may extend to more practical schemes, such as iterative decoding of the Low Density Parity-Check (LDPC) codes. 
Possible avenues of future research further include the performance analysis of the rare-transition regime in the absence of side information at the receiver.
Our conjecture is that in the limiting case, the channel sojourn time in each state is long enough for the receiver to estimate the state. 
For instance, for a channel with high correlation, patterns of errors with the same number of errors within a block are not equally likely.
In fact, the system is more prone to burst of errors when the channel quality is poor.
In other words, long channel memory enables the receiver to predict the channel quality.
Hence, one might expect similar performance when the state information is not provided at the receiver in the rare-transition regime.

\section{Appendix}

\subsection{Proof of Lemma~\ref{lemma:WeakConvergence}}
\label{sub:ctmc}

Distributions of the occupancy times for two-state discrete-time and continuous-time Markov chains have been studied previously. These distributions can be derived using bivariate generating functions and two-dimensional Laplace transforms, respectively~\cite{Pedler-jap71}.
Herein, we show how to adapt these approaches to derive the conditional distributions needed in our work.

The matrix of two-dimensional Laplace transforms for the distribution of the time spent in the first state over the time interval $[0,1]$ is given by
\begin{equation*}
\left[-\left(\mathbf{Q} - \begin{bmatrix}
\theta & 0\\ 0 & 0
\end{bmatrix}
-\phi \mathbf{I}\right)\right]^{-1} .
\end{equation*}
For example, the first entry in the matrix is equal to
\begin{equation*}
\frac{1}{u} + \frac{\mu\xi}{u(uv - \mu\xi)}
\end{equation*}
where $u = \phi+\theta+\mu$ and $v=\phi+\xi$.
The inverse two-dimensional Laplace transform of this entry gives the conditional distribution $f_{\eta_{1},S_{\mathrm{f}}|S_{\mathrm{i}}}(\cdot,1|1)$.
After this step, Lemma~2 in \cite{Pedler-jap71} can be employed to get the desired format in terms of modified Bessel functions.
These are the expressions presented in Lemma~\ref{lemma:WeakConvergence}.

\subsection{Proof of Lemma~\ref{thm:1}}
\label{sub:dtmc}

Let $a$ and $b$ be the numbers of transitions into and out of the initial state, respectively.
Then, we can write $c=a+b$ to denote the total number of transitions that occur up to time $N$.
From~\cite{Pedler-jap71}, we gather that
\begin{equation*}
\begin{split}
% SINGLE
P_{N_1|S_0}(m|1) &= (1-\alpha)^m (1-\beta)^{N-m}
\sum_{c=0}^{c_1} \binom{m}{a} \binom{N - m - 1}{b - 1}
\left(\frac{\alpha}{1-\beta}\right)^b \left(\frac{\beta}{1-\alpha}\right)^a
\end{split}
\end{equation*}
where
\begin{equation*}
c_{1}=\begin{cases}
N+\frac{1}{2}-\left|2m-\frac{1}{2}+N\right|, & m<N\\
0, & m=N.
\end{cases}
\end{equation*}
We can split the summation into two parts, one for odd and one for even values of $c$.
If $c=2k$, then $a=b=k$, and the corresponding sum represents $P_{N_1,S_N|S_0}(m,1|1)$.
If $c=2k+1$, then $a=k$, $b=k+1$, and the resulting sum is $P_{N_1,S_N|S_0}(m,2|1)$.
As such, we can write
\begin{equation*}
\begin{split}
&P_{N_1|S_0}(m|1)= (1 - \alpha)^{m}(1 - \beta)^{N-m} \\
&\times \Bigg( \sum_{k=0}^\infty \binom{m}{k} \binom{N - m - 1}{k - 1} \left(\frac{\alpha}{1 - \beta}\right)^{k} \left(\frac{\beta}{1- \alpha}\right)^{k}\\
&\quad + \sum_{k=0}^\infty \binom{m}{k} \binom{N - m - 1}{k} \left(\frac{\alpha}{1- \beta}\right)^{k+1}
\left(\frac{\beta}{1- \alpha}\right)^{k} \Bigg) .
\end{split}
\end{equation*}

We can set the upper and lower limits on $k$ to $0$ and $\infty$, respectively, since all other terms are automatically zero.
From the definition of ${}_2F_1 (-N + m + 1,-m;1;\lambda)$ in \cite[Lem.~1, pp.~383]{Pedler-jap71}, we see that
\begin{equation*}
\begin{split}
\left(\frac{\alpha}{1 - \beta}\right)
&\sum_{k=0}^{\infty} \binom{m}{k} \binom{N - m - 1}{k}
\left(\frac{\alpha}{1 - \beta}\right)^k
\left(\frac{\beta}{1 - \alpha}\right)^k \\
&=\left(\frac{\alpha}{1 - \beta}\right) {}_2F_1 (-N + m + 1,-m;1;\lambda).
\end{split}
\end{equation*}
Collecting these results, we obtain
\begin{equation*}
\begin{split}
P_{N_1,S_N|S_0}(m,2|1)
&= (1 - \alpha)^{m}(1 - \beta)^{N-m}
\left(\frac{\alpha}{1 - \beta}\right) \\
&\quad \times {}_2F_1 (-N + m + 1,-m;1;\lambda)
\end{split}
\end{equation*}
for $m=1,2,\ldots,N-1$. Clearly, for $m=0$ and $m=N$, this conditional probability is equal to zero.
Leveraging \cite{Pedler-jap71} and observing that
%$P_{N_1|S_0}(m|1) = P_{N_1,S_N|S_0}(m,1|1) + P_{N_1,S_N|S_0}(m,2|1)$,
\begin{equation*}
P_{N_1|S_0}(m|1) = P_{N_1,S_N|S_0}(m,1|1)+P_{N_1,S_N|S_0}(m,2|1),
\end{equation*}
we can write the simplified equation
\begin{equation*}
\begin{split}
&P_{N_1,S_N|S_0}(m,1|1) = (1-\alpha)^{m}(1-\beta)^{N-m} \\
&\times \! \left( {}_2F_1(-N + m,-m;1;\lambda)-{}_2F_1(-N + m + 1,-m;1;\lambda) \right)
\end{split}
\end{equation*}
for $m=1,2,\ldots,N-1$.
Moreover, $P_{N_1,S_N|S_0}(0,1|1)=0$ and $P_{N_1,S_N|S_0}(N,1|1)=(1-\alpha)^{N}$.
The remaining conditional probabilities can be derived in the same manner.

\bibliographystyle{IEEEtran}
\bibliography{WCLabrv,RareTransition}

\end{document}

%% file: Figures/bound_plots_avgres.tex
% This file was created by matlab2tikz v0.1.4.
% Copyright (c) 2008--2011, Nico Schlömer <nico.schloemer@gmail.com>
% All rights reserved.
%
% The latest updates can be retrieved from
%   http://www.mathworks.com/matlabcentral/fileexchange/22022-matlab2tikz
% where you can also make suggestions and rate matlab2tikz.
%
\begin{tikzpicture}[scale=0.75]

\begin{semilogyaxis}[%
scale only axis,
width=3.5in,
height=2.5in,
xmin=0.25, xmax=0.8,
ymin=1e-005, ymax=1,
yminorticks=true,
xmajorgrids,ymajorgrids,
xlabel={Code Rate ($\tilde{R}$)},
ylabel={Failure Probability},
axis on top,
legend entries={{Gallager-type Bound} $N=50$,{Gallager-type Bound} $N=75$,{Gallager-type Bound} $N=100$,{Rare - transition} $N=50$,{Rare - transition} $N=75$,{Rare - transition} $N=100$},
legend style={at={(1,0)},anchor={south east}}]
\addplot [
color=black,
dashed,
line width=1.0pt,
mark=o,
mark options={solid}
]
coordinates{ (0.25,0.000624627)(0.3,0.00290582)(0.35,0.0105345)(0.4,0.0310782)(0.45,0.076987)(0.5,0.163529)(0.55,0.302026)(0.6,0.489344)(0.65,0.698883)(0.7,0.881017)(0.75,0.981398)
  %(0.25,0.00162863)(0.3,0.00639656)(0.35,0.0197316)(0.4,0.0502905)(0.45,0.109501)(0.5,0.208259)(0.55,0.351096)(0.6,0.529528)(0.65,0.71803)(0.7,0.876425)(0.75,0.969947v)

};

\addplot [
color=black,
dashed,
line width=1.0pt,
mark=asterisk,
mark options={solid}
]
coordinates{
 (0.25,8.13177e-005)(0.3,0.000614999)(0.35,0.00324139)(0.4,0.0128583)(0.45,0.0403769)(0.5,0.103789)(0.55,0.223285)(0.6,0.407658)(0.65,0.636352)(0.7,0.850843)(0.75,0.975434)
};

\addplot [
color=black,
dashed,
line width=1.0pt,
mark=triangle,
mark options={solid,,rotate=180}
]
coordinates{
 (0.25,1.35385e-005)(0.3,0.000163962)(0.35,0.00122813)(0.4,0.00635319)(0.45,0.0244339)(0.5,0.0734126)(0.55,0.178083)(0.6,0.35614)(0.65,0.594018)(0.7,0.829353)(0.75,0.971017)
};

%\addplot [
%color=black,
%solid,
%line width=1.0pt,
%mark=o,
%mark options={solid}
%]
%coordinates{
% (0.25,0.000792121)(0.3,0.00353258)(0.35,0.0122949)(0.4,0.0349733)(0.45,0.0839249)(0.5,0.173525)(0.55,0.313498)(0.6,0.499286)(0.65,0.704336)(0.7,0.881136)(0.75,0.979541)
%};
%
%\addplot [
%color=black,
%solid,
%line width=1.0pt,
%mark=asterisk,
%mark options={solid}
%]
%coordinates{
% (0.25,0.000100155)(0.3,0.000732905)(0.35,0.00373933)(0.4,0.0143839)(0.45,0.0439147)(0.5,0.11012)(0.55,0.231987)(0.6,0.416473)(0.65,0.642044)(0.7,0.851627)(0.75,0.973924)
%};
%
%\addplot [
%color=black,
%solid,
%line width=1.0pt,
%mark=triangle,
%mark options={solid,,rotate=180}
%]
%coordinates{
% (0.25,1.62288e-005)(0.3,0.000191721)(0.35,0.00139895)(0.4,0.00705018)(0.45,0.0264491)(0.5,0.0776962)(0.55,0.184833)(0.6,0.363783)(0.65,0.599496)(0.7,0.830454)(0.75,0.969766)
%};

\addplot [
color=black,
solid,
line width=1.0pt,
mark=o,
mark options={solid,,rotate=180}
]
coordinates{
 (0.25,8.023e-004)(0.3,0.0035)(0.35,0.0123)(0.4,0.0349)(0.45,0.0815)(0.5,0.1618)(0.55,0.2809)(0.6,0.4332)(0.65,0.6009)(0.7,0.7582)(0.75,0.8808)
};

\addplot [
color=black,
solid,
line width=1.0pt,
mark=asterisk,
mark options={solid,,rotate=180}
]
coordinates{
 (0.25,9.9869e-005)(0.3,7.3144e-004)(0.35,0.0037)(0.4,0.0144)(0.45,0.0429)(0.5,0.1013)(0.55,0.2008)(0.6,0.3435)(0.65,0.5168)(0.7,0.6935)(0.75,0.8419)
};

\addplot [
color=black,
solid,
line width=1.0pt,
mark=triangle,
mark options={solid,,rotate=180}
]
coordinates{
 (0.25,1.62288e-005)(0.3,0.000191721)(0.35,0.00139895)(0.4,0.0070)(0.45,0.0261)(0.5,0.0709)(0.55,0.1557)(0.6,0.2879)(0.65,0.4599)(0.7,0.6461)(0.75,0.8111)
};
\end{semilogyaxis}

\end{tikzpicture}

%% file: Figures/boundvsexacts_plots_avgres.tex
% This file was created by matlab2tikz v0.1.4.
% Copyright (c) 2008--2011, Nico Schlömer <nico.schloemer@gmail.com>
% All rights reserved.
%
% The latest updates can be retrieved from
%   http://www.mathworks.com/matlabcentral/fileexchange/22022-matlab2tikz
% where you can also make suggestions and rate matlab2tikz.
%
\begin{tikzpicture}[scale=0.75]

\begin{semilogyaxis}[%
scale only axis,
width=3.5in,
height=2.5in,
xmin=0.25, xmax=0.8,
ymin=1e-006, ymax=1,
yminorticks=true,
xmajorgrids,ymajorgrids,
xlabel={Code Rate ($\tilde{R}$)},
ylabel={Failure Probability},
axis on top,
legend entries={$\text{ML }N\,=\,50$,$\text{ML }N\,=\,75$,$\text{MD }N\,=\,50$,$\text{MD }N\,=\,75$,$\text{Rare - transition }N\,=\,50$,$\text{Rare - transition }N\,=\,75$},
legend style={at={(1,0)},anchor={south east}}]
\addplot [
color=black,
solid,
line width=1.0pt,
mark=o,
mark options={solid}
]
coordinates{
 (0.25,0.000178344)(0.3,0.00067079)(0.35,0.00214493)(0.4,0.00591513)(0.45,0.0143075)(0.5,0.0308196)(0.55,0.0601084)(0.6,0.107283)(0.65,0.175714)(0.7,0.267464)(0.75,0.383104)
};

\addplot [
color=black,
solid,
line width=1.0pt
]
coordinates{
 (0.25,6.89775e-006)(0.3,4.99e-005)(0.35,0.000276952)(0.4,0.00120856)(0.45,0.00428744)(0.5,0.0127238)(0.55,0.0323583)(0.6,0.0716219)(0.65,0.139908)(0.7,0.243539)(0.75,0.380866)
};

\addplot [
color=black,
dashed,
line width=1.0pt,
mark=o,
mark options={solid}
]
coordinates{
 (0.25,0.000314153)(0.3,0.00115953)(0.35,0.00365273)(0.4,0.00998224)(0.45,0.0235163)(0.5,0.0480992)(0.55,0.087804)(0.6,0.149932)(0.65,0.2495)(0.7,0.344088)(0.75,0.445695)
};

\addplot [
color=black,
dashed,
line width=1.0pt
]
coordinates{
 (0.25,1.53958e-005)(0.3,0.000110869)(0.35,0.000609978)(0.4,0.00273185)(0.45,0.00855393)(0.5,0.0250595)(0.55,0.0599197)(0.6,0.115746)(0.65,0.203125)(0.7,0.322932)(0.75,0.469847)
};

\addplot [
color=black,
dash pattern=on 1pt off 3pt on 3pt off 3pt,
line width=2.0pt,
mark=o,
mark options={solid}
]
coordinates{
% (0.25,0.0018571)(0.3,0.0071387)(0.35,0.0215592)(0.4,0.0538627)(0.45,0.11517)(0.5,0.215584)(0.55,0.358613)(0.6,0.535127)(0.65,0.71997)(0.7,0.874625)(0.75,0.967494)
 (0.25,0.0018571)(0.3,0.0071387)(0.35,0.0215592)(0.4,0.0538627)(0.45,0.1113)(0.5,0.1953)(0.55,0.3074)(0.6,0.4420)(0.65,0.5873)(0.7,0.7275)(0.75,0.8464)
};

\addplot [
color=black,
dash pattern=on 1pt off 3pt on 3pt off 3pt,
line width=2.0pt
]
coordinates{
 %(0.25,0.000100155)(0.3,0.000732905)(0.35,0.00373933)(0.4,0.0143839)(0.45,0.0439147)(0.5,0.11012)(0.55,0.231987)(0.6,0.416473)(0.65,0.642044)(0.7,0.851627)(0.75,0.973924)
  (0.25,0.000100155)(0.3,0.000732905)(0.35,0.00373933)(0.4,0.0143839)(0.45,0.0429)(0.5,0.1013)(0.55,0.2008)(0.6,0.3435)(0.65,0.5168)(0.7,0.6935)(0.75,0.8419)
};

\end{semilogyaxis}

\end{tikzpicture}

%% file: Figures/bound2.tex
% This file was created by matlab2tikz v0.2.1.
% Copyright (c) 2008--2012, Nico Schlömer <nico.schloemer@gmail.com>
% All rights reserved.
%
% The latest updates can be retrieved from
%   http://www.mathworks.com/matlabcentral/fileexchange/22022-matlab2tikz
% where you can also make suggestions and rate matlab2tikz.
%
%
%
\begin{tikzpicture}[font=\large]

\begin{axis}[%
view={0}{90},
width=4.24791666666667in,
height=2.98958333333333in,
scale only axis,
xmin=0.25, xmax=0.75,
xlabel={Code Rate ($\tilde{R}$)},
xmajorgrids,
ymin=0, ymax=1,
ylabel={Bound on Tail Probability, ${\Pr} (\tilde{Q} > 5)$},
ymajorgrids,
axis lines=left,
title={Approximate Bound on Probability of Buffer Overflow},
legend style={nodes=right},
legend style={at={(1,0)},anchor={south east}}]

\addplot [
color=black,
solid,
line width=1.0pt,
mark=o,
mark options={solid}
]
coordinates{
 (0.75,1)(0.7,1)(0.65,1)(0.6,1)(0.55,1)(0.5,1)(0.45,1)(0.4,0.252629165822919)(0.35,0.212853598489266)(0.3,0.286852988037856)(0.25,0.747778031912185)
};

\addlegendentry{$N$ = 75};

\addplot [
color=black,
solid,
line width=1.0pt,
mark=square,
mark options={solid}
]
coordinates{
 (0.75,1)(0.7,1)(0.65,1)(0.6,1)(0.55,1)(0.5,0.137210534453183)(0.45,0.077799668742969)(0.4,0.101031602442602)(0.35,0.203682421267488)(0.3,0.403420247018975)(0.25,0.719678311503304)
};

\addlegendentry{$N$ = 125};

\addplot [
color=black,
solid,
line width=1.0pt,
mark=asterisk,
mark options={solid}
]
coordinates{
 (0.75,1)(0.7,1)(0.65,1)(0.6,1)(0.55,0.155634849293506)(0.5,0.0657542884431945)(0.45,0.0820597066929997)(0.4,0.138425564643363)(0.35,0.260141050367492)(0.3,0.545669916457175)(0.25,1)
};

\addlegendentry{$N$ = 170};

\addplot [
color=black,
solid,
line width=1.0pt,
mark=star,
mark options={solid}
]
coordinates{
 (0.75,1)(0.7,1)(0.65,1)(0.6,0.51105915520834)(0.55,0.0824387962421791)(0.5,0.0792933070239446)(0.45,0.120704650410619)(0.4,0.204632885722539)(0.35,0.348445293049352)(0.3,0.706411149695612)(0.25,1)
};

\addlegendentry{$N$ = 225};

\addplot [
color=black,
solid,
line width=1.0pt,
mark=triangle,
mark options={solid}
]
coordinates{
 (0.75,1)(0.7,1)(0.65,1)(0.6,0.241651474520346)(0.55,0.093948737144523)(0.5,0.118332550927866)(0.45,0.175203763944271)(0.4,0.292055279834126)(0.35,0.499072170407217)(0.3,1)(0.25,1)
};

\addlegendentry{$N$ = 275};

\addplot [
color=black,
solid,
line width=1.0pt,
mark=diamond,
mark options={solid}
]
coordinates{
 (0.75,1)(0.7,1)(0.65,1)(0.6,0.218667740588482)(0.55,0.136973485725367)(0.5,0.180515099034899)(0.45,0.274005111147662)(0.4,0.428970074933491)(0.35,0.722021675824395)(0.3,1)(0.25,1)
};

\addlegendentry{$N$ = 325};

\addplot [
color=black,
solid,
line width=1.0pt,
mark=triangle,
mark options={solid,,rotate=90}
]
coordinates{
 (0.75,1)(0.7,1)(0.65,1)(0.6,0.255362728002582)(0.55,0.220533259593)(0.5,0.294595592166901)(0.45,0.422348291058601)(0.4,0.645850871609857)(0.35,1)(0.3,1)(0.25,1)
};

\addlegendentry{$N$ = 375};

\end{axis}
\end{tikzpicture}

%% file: Figures/exact.tex
% This file was created by matlab2tikz v0.2.1.
% Copyright (c) 2008--2012, Nico Schlömer <nico.schloemer@gmail.com>
% All rights reserved.
%
% The latest updates can be retrieved from
%   http://www.mathworks.com/matlabcentral/fileexchange/22022-matlab2tikz
% where you can also make suggestions and rate matlab2tikz.
%
%
%
\begin{tikzpicture}[font=\large]

\begin{axis}[%
view={0}{90},
width=4.24791666666667in,
height=2.98958333333333in,
scale only axis,
xmin=0.25, xmax=0.75,
xlabel={Code Rate ($\tilde{R}$)},
xmajorgrids,
ymin=0, ymax=1,
ylabel={Tail Probability of Transmit Buffer, $\Pr (Q > 5)$ },
ymajorgrids,
axis lines=left,
title={Buffer Overflow Probability},
legend style={nodes=right},
legend style={at={(1,0)},anchor={south east}}]
\addplot [
color=black,
solid,
line width=1.0pt,
mark=o,
mark options={solid}
]
coordinates{
 (0.75,1)(0.7,1)(0.65,1)(0.6,1)(0.55,1)(0.5,0.260554327846649)(0.45,0.108745014980012)(0.4,0.0842012852645387)(0.35,0.155901377865541)(0.3,0.273446330958796)(0.25,0.746078107011096)
};
\addlegendentry{$N$ = 75};
\addplot [
color=black,
solid,
line width=1.0pt,
mark=square,
mark options={solid}
]
coordinates{
 (0.75,1)(0.7,1)(0.65,1)(0.6,1)(0.55,0.274922617804184)(0.5,0.0549659113268103)(0.45,0.0613738996988479)(0.4,0.0978904631597134)(0.35,0.203341396841889)(0.3,0.403408319939864)(0.25,0.719680512569038)
};
\addlegendentry{$N$ = 125};

\addplot [
color=black,
solid,
line width=1.0pt,
mark=asterisk,
mark options={solid}
]
coordinates{
 (0.75,1)(0.7,1)(0.65,1)(0.6,1)(0.55,0.0743453460822128)(0.5,0.0528228675840124)(0.45,0.0799554901538792)(0.4,0.138283419741642)(0.35,0.260136249928191)(0.3,0.545669671147412)(0.25,1)
};
\addlegendentry{$N$ = 170};
\addplot [
color=black,
solid,
line width=1.0pt,
mark=star,
mark options={solid}
]
coordinates{
 (0.75,1)(0.7,1)(0.65,1)(0.6,0.407819815838069)(0.55,0.0630270232943787)(0.5,0.0761847641657128)(0.45,0.120540469263419)(0.4,0.204628691479067)(0.35,0.348445168107902)(0.3,0.70641114792933)(0.25,1)
};
\addlegendentry{$N$ = 225};
\end{axis}
\end{tikzpicture}